\documentclass[preprint,showpacs,preprintnumbers,amsmath,amssymb,showkeys]{revtex4}
\usepackage{graphicx}
\usepackage{dcolumn}
\usepackage{bm}
\usepackage{placeins}
\usepackage{subfigure}
\begin{document}
\title{Cellular polarization: interaction between extrinsic bounded noises and  wave-pinning mechanism}
\author{Sebastiano de Franciscis}
\author{Alberto d'Onofrio (Corresponding author)}
\email{alberto.donofrio@ieo.eu}
\affiliation{
European Institute of Oncology, Department of Experimental Oncology,Via Ripamonti 435, I20141 Milano (Italy)}%
\date{\today}
\begin{abstract} Cued and un-cued cell polarization is a fundamental mechanism in cell biology. As an alternative to the classical Turing bifurcation, it has been proposed that the cell polarity might onset by means of the well-known phenomenon of wave-pinning (Gamba et al, PNAS, 2005). A particularly simple and elegant model of wave-pinning has been proposed by Edelstein-Keshet and coworkers (Biop. J., 2008).\\
However, biomolecular networks do communicate with other networks as well as with the external world. As such, their dynamics has to be considered as perturbed by extrinsic noises. These noises may have both a spatial and a temporal correlation, but any case they must be bounded to preserve the biological meaningfulness of the perturbed parameters.\\
Here we numerically show that the inclusion of external spatio-temporal bounded perturbations may sometime destroy the polarized state. The polarization loss depends on both the extent of temporal and spatial correlations, and on the kind of adopted noise.\\ Namely, independently of the specific model of noise, an increase of the spatial correlation induces an increase of the probability of polarization. However, if the noise is spatially homogeneous then the polarization is lost in the majority of cases.  \\
On the contrary, an increase of the temporal autocorrelation of the noise induces an effect that depends on the noise model.
\end{abstract}
\maketitle
\section{Introduction} 
The formation of two distinct spatial domains able to distinct in a single cell two separated parts, let us say a "tail" and a "head", is called cellular polarization \cite{OnsumRao09}. This phenomenon is at the basis of two fundamental bioprocesses: the asymmetric cell division and the chemotactic motion of some kinds of cells (for example neutrophils) \cite{OnsumRao09}.

The first is a patterning that has to regularly repeat until the replicating potential of a proliferating cell is exhausted (theoretically never for stem cells), thus it is thought to be essentially \emph{un-cued} \cite{sohrmann2003pol}. However, there are models, such as that by Ortoleva and Ross \cite{OrtolevaRoss73}, that postulate a role of random cues also for the onset of asymmetric division. Finally, also in absence of mitosis, a cell can experience symmetry-breaking and start moving in random directions \cite{pmsb}.

On the contrary, the patterning during chemotactic motion induced by an external cue: the gradient of the chemoattractant to be followed (or of the chemorepellor to be avoided) \cite{murray,devr}.

In both cases it is necessary that a pattern "head"-"tail" must be formed. From the biophysical point of view, both cases are classified as pattern onset in non-equilibrium systems. Patterns formation in biosystems is a fundamental topic of computational biology, and it is largely influential in experimental  biology. This area of investigation has been started by Alan Turing's very well-known paper on morphogenesis \cite{Tur52}, where he modeled the onset of a pattern in two kinds of multi-cellular structures (a ring and a sphere) as a symmetry-breaking bifurcation driven by a strong difference in the diffusion of two morphogens. This mechanism is called the Turing's bifurcation. In early seventies the Turing mechanism has been biologically substantiated by Gierer and Meinhardt \cite{GierMein72, Mein82, Mein99}, who introduced two possible patterning mechanisms based on Turing bifurcation. The first is the Activation-Inhibitor model, where the pattern is induced by the reaction-diffusion interplay between a \emph{short range} self-activator (which is membrane-bound or, any case, has a very low diffusion coefficient) and a \emph{long-range} inhibitor (which has a large diffusion coefficient). The second model is the Activation-Depletion model, where two proteins interact: the first is again a short-range self-activator, but the second is no more an inhibitor. On the contrary, the second protein is, instead, depleted by the activator.

These pioneering studies generated a wide amount of literature, which is increasingly linked with experimental data \cite{Meinhardt2012,Maini2012,sohrmann2003pol,Strier07, Mogilner_Allard_Wollman_2012}.

In 2005, Gamba and colleagues \cite{Gamba2005, GambaPRL07, GambaJSM09, GambaPLOS12} proposed a stochastic model of chemotaxis-induced spatial symmetry breaking in a single cell. The central core of this model was the interplay between two membrane-bound molecules and two cytosolic molecules. The presence of some feedbacks (with related bistability)  and the difference of diffusion coefficients induced a mechanism of phase separation that is different from the ones based on Turing's bifurcation. Quite interestingly, in the supplementary materials of \cite{Gamba2005} Gamba et al. proposed a simplified and mean-field version of their model that exhibits, in one spatial dimension, a traveling wave whose velocity decreases until the front stops.
 
In 2008, Edelstein-Keshet and colleagues proposed in \cite{k2008,Keshet2012,KeshetPLOS11,KeshetSIAM11} a similar but simpler biomechanism leading to the onset of single-cell polarization due to the 'freezing' of an initially traveling wave of large concentration of a membrane-bound protein.
Indeed, this elegant and minimalist mechanism is based on the inter-conversion of a membrane-bound active protein 'A' in its cytosolic inactive form 'B', where 'A' positively feedbacks on its activation. Also this process induces the initial onset of a wave-front that stops. This phenomenon is called in physics 'wave-pinning'. The wave-pinning-induced polarization will be here called WPP.

The WPP process is biologically back-grounded, since it well fits to some key proteins involved in cell polarization \cite{k2008}, such as Rho-GTPases, which switch between active membrane-bound and inactive cytosolic forms.

Some similarities exists between the WPP and the Activation-Depletion induced polarization (ADP) in that: \textit{i)} in WPP the two involved molecular forms have very different diffusion coefficients; and \textit{ii)} the positive feedback of 'A' on its activation that depletes 'B' may remember the depletion mechanism. However, in WPP the patterning mechanism is totally different, since in it no Turing bifurcation is observed, and, conversely, in the Turing-based models no waves are observed. Moreover, one of the 'essential ingredients' \cite{k2008} of WPP model is the presence of bistability (see also \cite{Gamba2005, GambaJSM09}), making it a genuinely non-linear model, whereas Turing bifurcation stems from the linearization of reaction-diffusion equations. Finally, in WPP the total mass of the two forms 'A' and 'B' is conserved.

Summarizing, from a theoretical point of view,  although they are two quite different mechanisms, both ADP and WPP have equivalent effects of inducing a pattern. However, passing from the general principles to the concrete numerical simulations in case of realistic values of the parameters, in \cite{k2008} it has been shown that the WPP has (at least for the examined range of parametric values) an important advantage on ADP: it is extremely rapid. Indeed, in WPP the pre-patterning transient has a characteristic time of more or less $10$ seconds, whereas ADP would require more than $10$ minutes \cite{k2008}. The latter time is too long and would make ADP unsuitable, especially for chemotaxis-driven polarization. 

This temporal difference in the duration of respective transients makes WPP a very attractive mechanism to be experimentally validated.

Moreover, from the point of view of theoretical physics, the WPP model is one of few examples (see also \cite{Sepulchre00}) of robust wave-pinning. Indeed, standard wave-pinning condition is fulfilled for isolated points of a key parameter, and also for very small parametric changes the wave restarts traveling. 

Very recently, it has been published a paper \cite{Keshet2012} on the robustness of the WPP model with respect to the intrinsic stochasticity, i.e. by considering the fact that the molecular distribution is discrete and not continuous. The main result of that investigation is that for very small number of molecules the wave collapses and the WPP is destroyed.

Here we want to study another an equally important problem: the robustness of WPP with respect to the unavoidable presence of extrinsic noise. 

As we mentioned above, one of the basis of WPP is the presence of multistability. The interplay between extrinsic noise and multistability is of fundamental relevance in chemistry and systems biology \cite{enb,cpsdmbn,GiulioDon12PONE12}, as shown since the pioneering work by Kramers \cite{Kramers1940284}. In a nonlinear system, indeed, the presence of noise may cause the transition from a state to another \cite{PhysRevA.40.5447} (and other important effects \cite{GiulioDon12PONE12}). This in the WPP case might mean that the noise could induce the switch from a polarized to a non-polar molecular distribution. Moreover, the presence of noise may deeply affect the dynamics of traveling waves, as reviewed in \cite{Panja200487}.

Moreover, the above-mentioned rapid duration of the transient leading to the full polarization suggests that the WPP might not be robust to realistic colored extrinsic noises representing the interplay of the WPP mini-network with other biomolecular networks that have the same or even larger characteristic times.

We investigate here this point by means of numerical simulations of the WPP model in the case that its parameters are perturbed by bounded colored spatio-temporal noises.
\section{Unbounded and bounded noises}
Let us consider the well-known zero-dimensional Ornstein-Uhlenbeck stochastic differential equation:
\begin{equation}\label{oue}
\xi^{\prime} (t)= -\frac{1}{\tau_c}\xi(t) + \frac{\sqrt {2 D} }{\tau_c}\eta(t),
\end{equation}
where $\tau_c>0$, $\sqrt {2 D}$  is the noise strength and $\eta(t)$ is a Gaussian white noise of unitary intensity$
\langle \eta(t)\eta(t_1)\rangle$ $= \delta(t-t_1)$.
It is well-known that solution of equation (\ref{oue}) is a Gaussian colored stochastic process with autocorrelation:$
\langle \xi(t)\xi(t_1)\rangle$  $\propto \ exp(- |t-t_1|/\tau_c)$. In \cite{GO92} equation (\ref{oue}) 
was generalized in a spatially extended setting by including in it the most known and simple spatial coupling, the Laplace operator, yielding the following partial differential Langevin equation:
\begin{equation}\label {gener}
\partial_t \xi (x,t)= D_{noise}\nabla^2 \xi(x,t) -\frac{1}{\tau_c}\xi(x,t) + \frac{\sqrt {2 D} }{\tau_c}\eta(x,t),
\end{equation}
where $D_{noise}>0$ is the spatial correlation strength \cite{GO92} of $\xi (x,t)$.

Two strategies have mainly been adopted so far for defining zero-dimensional bounded noises: i) nonlinear filtering unbounded noises; ii) applying bounded functions to unbounded noises. 

The first approach was employed in \cite{CaiLin, CaiLinII, caiwu}, where the following family of bounded noises was introduced:
\begin{equation}\label{Cai}
\dot{\xi} (t)= -\frac{1}{\tau_c}\xi(t) + \sqrt{\frac{B^2 -\xi^2}{\tau_c(1+z)}}\eta(t),
\end{equation}
where  $\eta(t)$ is a Gaussian white noise with $\langle \eta(t)\eta(t_1) \rangle= \delta(t-t_1)$. The bounded nature of the noise described in (\ref{Cai}) easily follows from the fact that $\xi=+B \ (-B)$ implies $\dot{\xi}<0 \ (> 0)$. The process $\xi(t)$ has zero mean and the same autocorrelation of the OU process \cite{CaiLin, CaiLinII}, and its stationary probability density is given by: $ P_{CL}(\xi)$ $ = A \left(B^2 - \xi^2\right)_{+}^{z}$. For $z>0$ the distribution is unimodal and centered in $0$, while for $-1<z<0$ it is bimodal, having a "horned" distribution with two vertical asymptotes at $\xi\rightarrow \pm B$. 

 In order to define spatio-temporal bounded noises based on  the Cai-Lin noises, in \cite{deFradOnpre} we adopted an approach analogous to the one employed in \cite{GO92pla,GO92} and in equation (\ref{gener}) to extend the OU process. Thus, the spatio-temporal Cai-Lin noise can be defined as follows:
\begin{equation}\label{sptCai}
\partial_t \xi (x,t)= D_{noise}\nabla^2 \xi(x,t) -\frac{1}{\tau_c}\xi(x,t) + \sqrt{\frac{B^2 -\xi^2}{\tau_c(1+z)}}\eta(x,t).
\end{equation}

The sine-Wiener noise \cite{bobryk} is obtained by applying the bounded function $h(u) = B \sin(\sqrt{2/\tau_c}u)$ to a random walk $W(t)$ defined as $W'= \eta(t)$, where $\eta(t)$ is a white noise of unitary intensity, yielding:
\begin{equation}
\zeta(t)= B \sin\left( \sqrt{ \frac{2}{\tau_c}}W(t) \right).
\end{equation}
The stationary probability density of $\zeta(t)$ is given by
$P_{SW}(\zeta) =$ $ 1/(\pi \sqrt{B^2 -\zeta^2})$,
thus: $P_{eq}(\pm B)=+\infty$. 

In \cite{deFranDOnoSW}, as a natural spatial extension of the sine-Wiener noise, we defined the following spatio-temporal noise:
\begin{equation}\label{XXX}
\zeta(x,t)= B \sin\left( 2 \pi \xi(x,t) \right),
\end{equation}
where $\xi(x,t)$ is defined by equation (\ref{gener}). 

Detailed studies of the properties of the above-defined spatio-temporal noises have been performed in \cite{deFradOnpre, deFranDOnoSW}.
\section{Background on the WPP model}
The WPP model describes the interplay of two different forms (active, 'A', and inactive, 'B') of a biomolecule, of which the active one is membrane-bound and the other is located in the cytosol and has a very large diffusion coefficient. Denoting with $a(x,t)$ and $b(x,t)$ the concentrations of, respectively, A and B, the WPP model reads :
\begin{align}
\partial_t a &= D_a \nabla^2 a + f(a,b; p) \label{wppa}\\
\partial_t b &= D_b \nabla^2 b - f(a,b; p), \label{wppb}
\end{align}
where $D_a$ and $D_b$ are the diffusion coefficients of A and B, and $p$ is a vector of parameters (see below). Since A is membrane-bound and B is free in the cytosol, the following constraint must hold: 
\begin{equation}\label{gg}D_b \gg D_a.\end{equation}

A central hypothesis in \cite{k2008} is that A positively feedbacks on its activation. Thus, it is assumed that  $f(a,b; p)$ is of the form $g(a; p)b$ with $\partial_a g(a)>0$. Namely, in \cite{k2008} it is employed the following functional form:
\begin{equation}\label{f}
f(a,b; p)=b \left( k_a + \frac{\gamma \omega a^2}{K + \omega a^2} \right) -\delta a.
\end{equation}
As a consequence, the vector of parameters reads
\begin{equation}
p =(k_a, \gamma,K,\omega,\delta).
\end{equation}
The boundary conditions are assumed of the no-flux type:
\begin{equation}
\nabla a(t,0) = \nabla a(t,L) = \nabla b(t,0) = \nabla b(t,L) =0, 
\end{equation}
which, of course, imply the mass conservation:
\begin{equation}
\int_0^L \left( a(t,x) + b(t,x) \right) dx = Q.
\end{equation}
Note that the fundamental assumption
$D_b \gg D_a$ ensures the correct working, and the robustness, of the wave-pinning mechanism.

In \cite{k2008},  Mori and co-workers showed that not only the above system is able to generate a polarization in response to initial transient cues and in absence of cues, but also that the onset of this pattern is faster than in the Turing mechanism. Indeed, they showed that the time to polarize is of the order of $1$ to $10$ seconds, and the effective time to 'complete the polarization' \cite{k2008} is of about $30$ seconds. 
\section{Including extrinsic noise in the WPP Model}
We phenomenologically take into the account the interplay with extrinsic noise by assuming, in the most general case, that:
\begin{align}
\partial_t a &= D_a \nabla^2 a + f(a,b; \widehat{p}(x,t) ) + k_S(x,t) b  \label{wppanoise}\\
\partial_t b &= D_b \nabla^2 b - f(a,b; \widehat{p}(x,t) ) - k_S(x,t)  b,\label{wppbnoise}
\end{align}
where $k_S(x,t) b$ is the initial transient cue, which is null after a short time, and:
\begin{equation}\label{pnoisy}
\widehat{p}_i(x,t)=p_i (1+\xi_i(x,t) )>0,
\end{equation}
where $\xi_i(x,t)$ is a bounded noise of the Cai-Lin or sine-Wiener type. We employ two different kinds of noises in line with the recent literature on bounded noises \cite{dongan, pre, deFradOnpre, deFranDOnoSW}, which in other context showed that the statistical characteristics of a system perturbed by a bounded stochastic process depends not only on the bound of the noise, but also on its finer structure.

Unfortunately, no analytical tools are currently available to investigate the effects of bounded spatio-temporal perturbations, so that we shall resort to numerical simulations. 

We consider two kind of numerical experiments corresponding to, respectively, cued and un-cued polarization of a cell:

\emph{i)} external graded transient cue, modeled by:
\begin{equation}\label{gcue1}
k_S(x,t)=s(t)(1-x/L) 
\end{equation}

\begin{equation}\label{gcue2}
s(t)      = \begin{cases}
S                                                     & t\in [0,t_1]      \\
S\left(1-\frac{t-t_1}{t_2-t_1}\right) & t\in [t_1,t_2]. \\
0 & T>t_2 \end{cases}
\end{equation}

\emph{ii)} random initial conditions with no external transient cue, i.e.:
\begin{equation}\label{rcue}
a(x,0)=R\mbox{ }\eta(x) a_{-}, \quad b(x,0)=2.0,
\end{equation}
where $a_{-}=0.2683312$ corresponds to the lower homogeneous steady state of WPP model and $\eta(x)$ is a spatial noise uniformly distributed in $[0,1]$. 

In both cases Mori \emph{et al.} \cite{k2008} showed that the transitory lasts less than $200$ seconds, at which time the transitory is well established. However, the presence of spatio-temporal correlations might induce slower transients, so that we simulated all systems up to time $T=600 \quad s$.
A first statistics we measured for each realization of our stochastic process is the following:
\begin{equation}\label{Xavg}
\langle x \rangle = \frac{\int_0^L x a(x,T)dx}{\int_0^L a(x,T)dx},
\end{equation}
which is the average 'position' $x$ weighted by the normalized version of $a(x,T)$. This statistics measures where the polarization is oriented. Of course, in the case of cued polarization this measure is mainly relevant to assess whether the polarization is in line with the cue, or if it is not present in the final simulation time. The second and more important statistics should then measure some variance of the density. One might be tempted to measure also this statistics in the final simulation time. However, during our numerical simulations we observed cases where the stochastically perturbed front temporarily became flat to then recover a polarized state, and so on. As a consequence, the analysis of the variability of the profile of $a$ must also consider the past history of the system. Namely, starting from the instantaneous amplitude of the distribution:
\begin{equation}
\delta_t = \left(\textrm{Max}_{x \in [0,L]} a(x,t)\right) - \left(\textrm{Max}_{x \in [0,L]} a(x,t)\right)
\end{equation}
we defined the following statistics:
\begin{equation}\label{deltahat}
\Delta_t = \textrm{min}_{t \in [200,T]}  \delta_t
\end{equation}
Note that in case of  noisy initial conditions, we expect an equiprobable 
distribution of the polarization, and also cases where both $\Delta_t$ is 'large' and $\langle x \rangle \approx 0.5$, which indicate the presence of a 'central hump' in the distribution of $a$.

As far as the characteristics of the employed noises are concerned, we shall assume that their temporal autocorrelation has a time-scale of $10$ seconds. Also far larger values would be acceptable, since there is a number of biomolecular processes that are characterized by long time-scales. However, in those cases the loss of polarization would be almost sure. As far as the spatial scales of the noises, we considered three cases: \emph{i)} spatially white noises; \emph{ii)} spatially uniform noises; and \emph{iii)} finite spatial correlation. 

\section{Numerical Simulations}
In line with \cite{k2008}, in all our simulations we set the following values for the parameters of the WPP model: $D_a=0.1 \mbox{ }\mu m^{2}s^{-1}$, $D_b=10 \mbox{ }\mu m^{2}s^{-1}$, $\omega=1$, $\delta=1\mbox{ }s^{-1}$, $\gamma=1\mbox{ }s^{-1}$, $k_a=0.067\mbox{ }s^{-1}$, $K=1$ and cell size $L=10 \mbox{ }\mu m$ \cite{k2008, Zhang98, Sako00, Postma04}.

In all our simulations, the stochastic predictor-corrector algorithm \cite{Sagues07} for It\^{o} partial stochastic differential equations was adopted to simulate the noise. Moreover, in case of the Cai-Lin noise the transformation described in \cite{deFradOnpre} was also adopted. The WPP model was then simulated by means of the second-order Runge-Kutta algorithm.

Finally, a remark: in next sections, for the sake of the simplicity, instead of writing phrases like "this simulation of the model suggests that the cell does something" we shall concisely write (with an abuse of meaning) "the cell does something". 

\subsection{External Cues}
In the first round of simulations we stimulated the system with the graded external cue, described in equations (\ref{gcue1})-(\ref{gcue2}), with $S=0.07 m \mbox{ }s^{-1}$, $t_{1}=20 s$ and $t_2=25 s$. The initial conditions were:
$(a(x,0),b(x,0))=(a_{-}, b_{0})$ $\approx$ $(0.2683, 20)$, corresponding to the lower homogeneous steady state of the unperturbed system \cite{k2008}. In absence of noise we obtained, as in \cite{k2008}, that the cell polarizes with time-scales of the order of $50 s$. 

By introducing an extrinsic noise in one of the relevant parameters of function $f(a,b)$ via formula (\ref{pnoisy}), we obtained a complex pattern of responses, in dependence of the spatial and temporal parameters of the noise, as well as of its type.

In some cases, which we are going to illustrate, our stochastic model predicts that the cell polarization is preserved, although, of course, the wave-front experiences some random fluctuations; in other cases the cell depolarizes. In these cases, the profile of the concentration of $A$ is flat and oscillating.

We never observed an inversion of the polarization, i.e. the polarized state follows in all cases the external cue.

\begin{figure}[b]
\begin{center}
\subfigure[]
{
\label{A}
\includegraphics[width=0.25\textwidth]{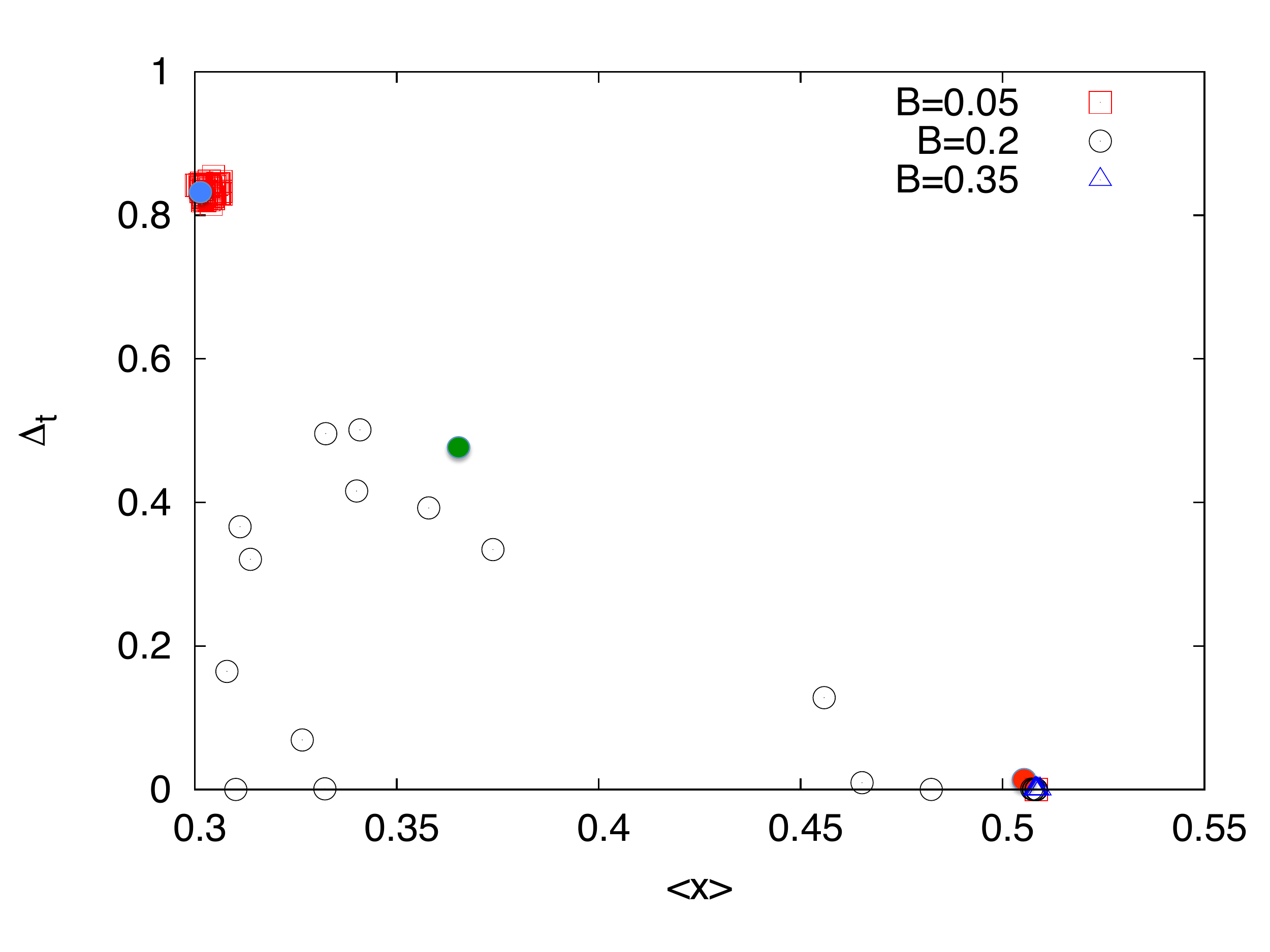}
}
\subfigure[]
{
\label{B}
\includegraphics[width=0.25\textwidth]{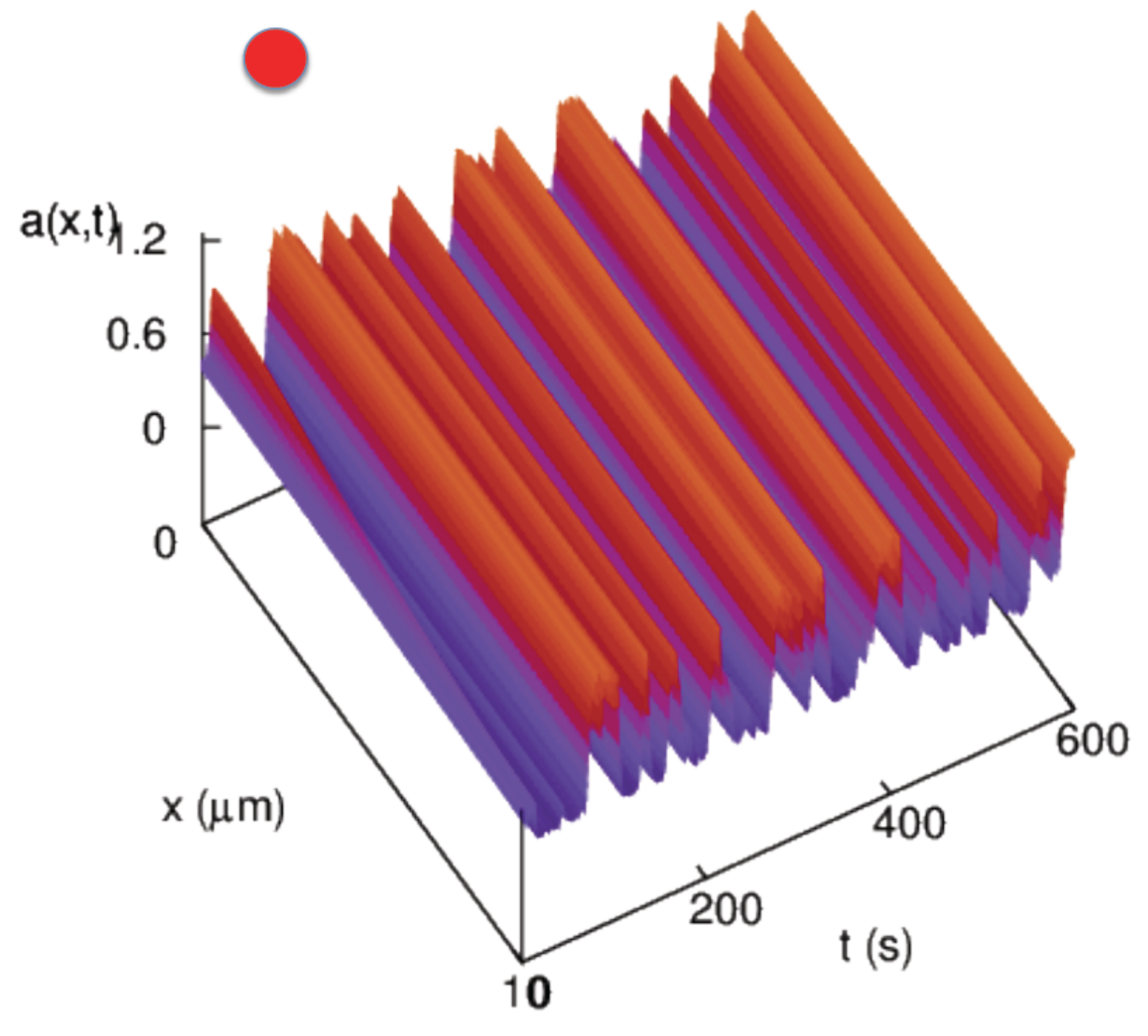}
}\\
\subfigure[]
{
\label{C}
\includegraphics[width=0.25\textwidth]{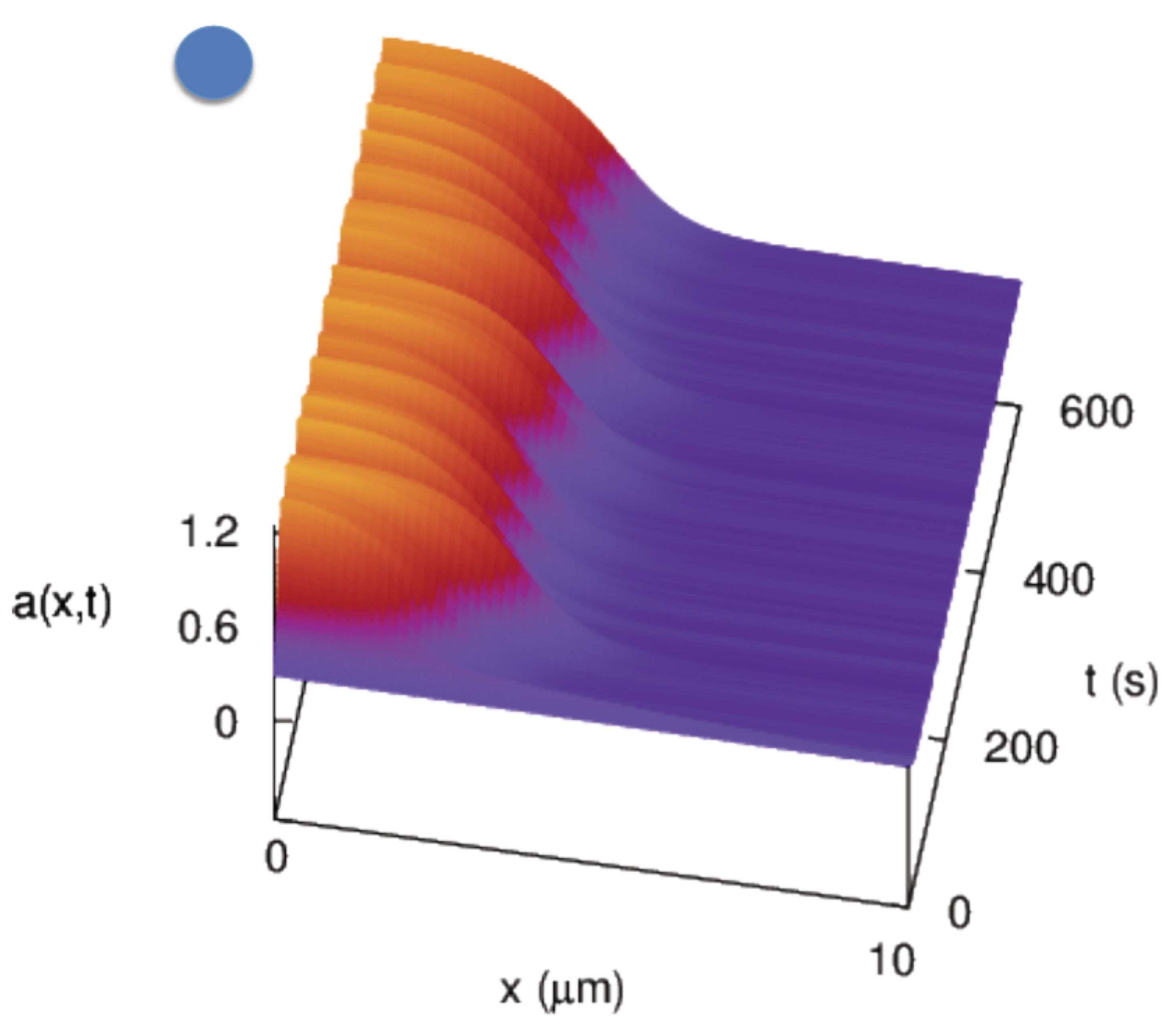}
}
\subfigure[]
{
\label{D}
\includegraphics[width=0.25\textwidth]{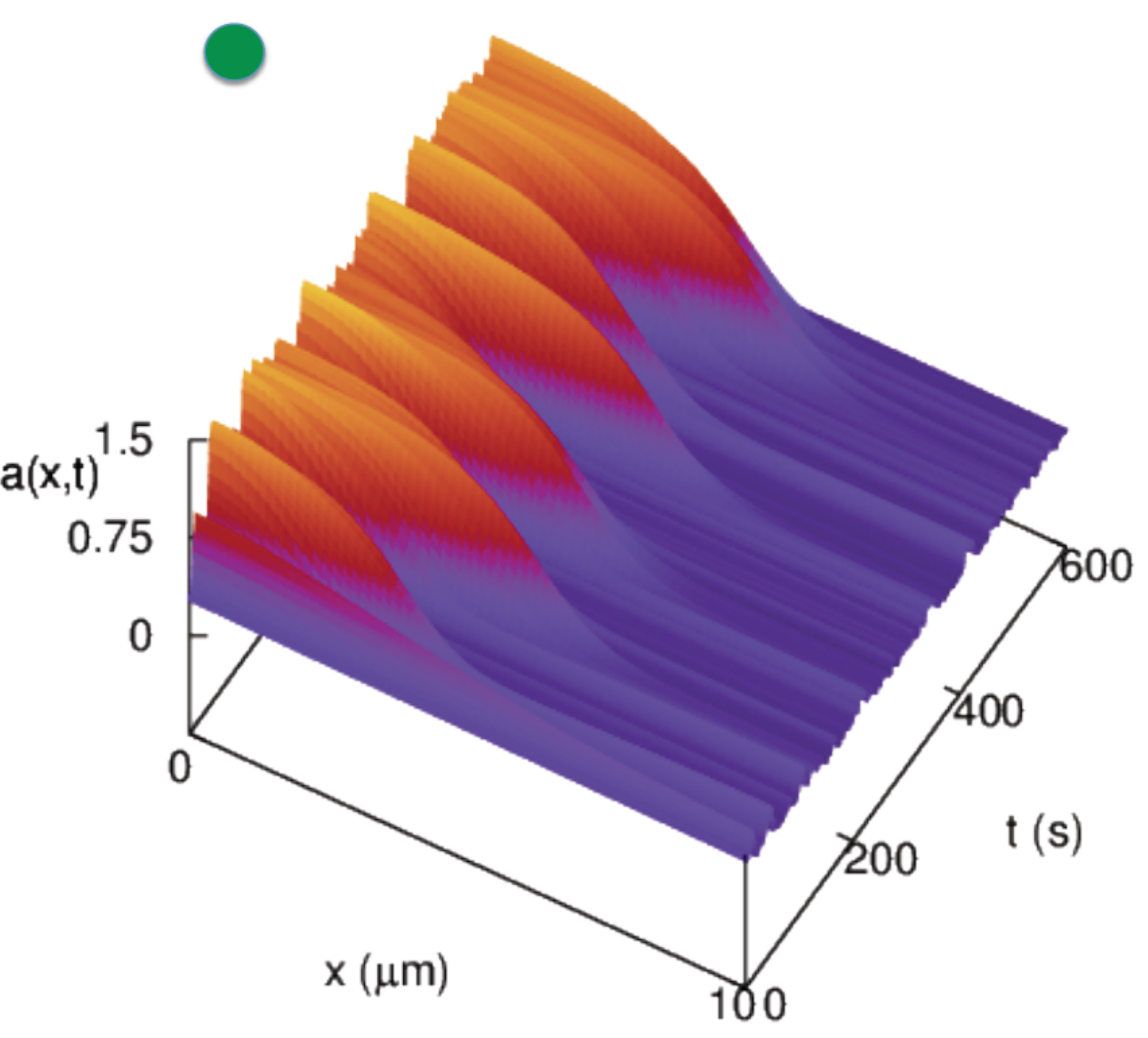}
}
\end{center}
\caption{Bounded stochastic perturbation of parameter $\delta$. Type of noise: spatially white Cai-Lin noise with temporal correlation $\tau_c=10 \; s$. Panel \subref{A}: scatterplot $(\Delta_t,\langle x \rangle)$ for the density $a(x,t)$.  For large and intermediate values of $\Delta_t$ the system is polarized, see panels \subref{C} and \subref{D}, and the fluctuations of the polarization front are more pronounced for decreasing $\Delta_t$. For $\Delta_t\simeq0$ the system is not polarized, and is characterized by a (spatially constant) temporal oscillation dynamic, see panel \subref{B}.
}
\label{fig_1}
\end{figure}

Typical configurations are shown in figure \ref{fig_1}, where both a scatterplot $(\Delta_t,\langle x \rangle)$ and three specific realizations are shown. 

We now illustrate the statistical response of the model to bounded perturbations that are spatially white, i.e.  $D_{noise}=0$.

Initially we shall consider perturbations in the parameter $\delta$.

\begin{figure}[t]
\begin{center}
\subfigure[]
{
\label{A}
\includegraphics[width=0.45\textwidth]{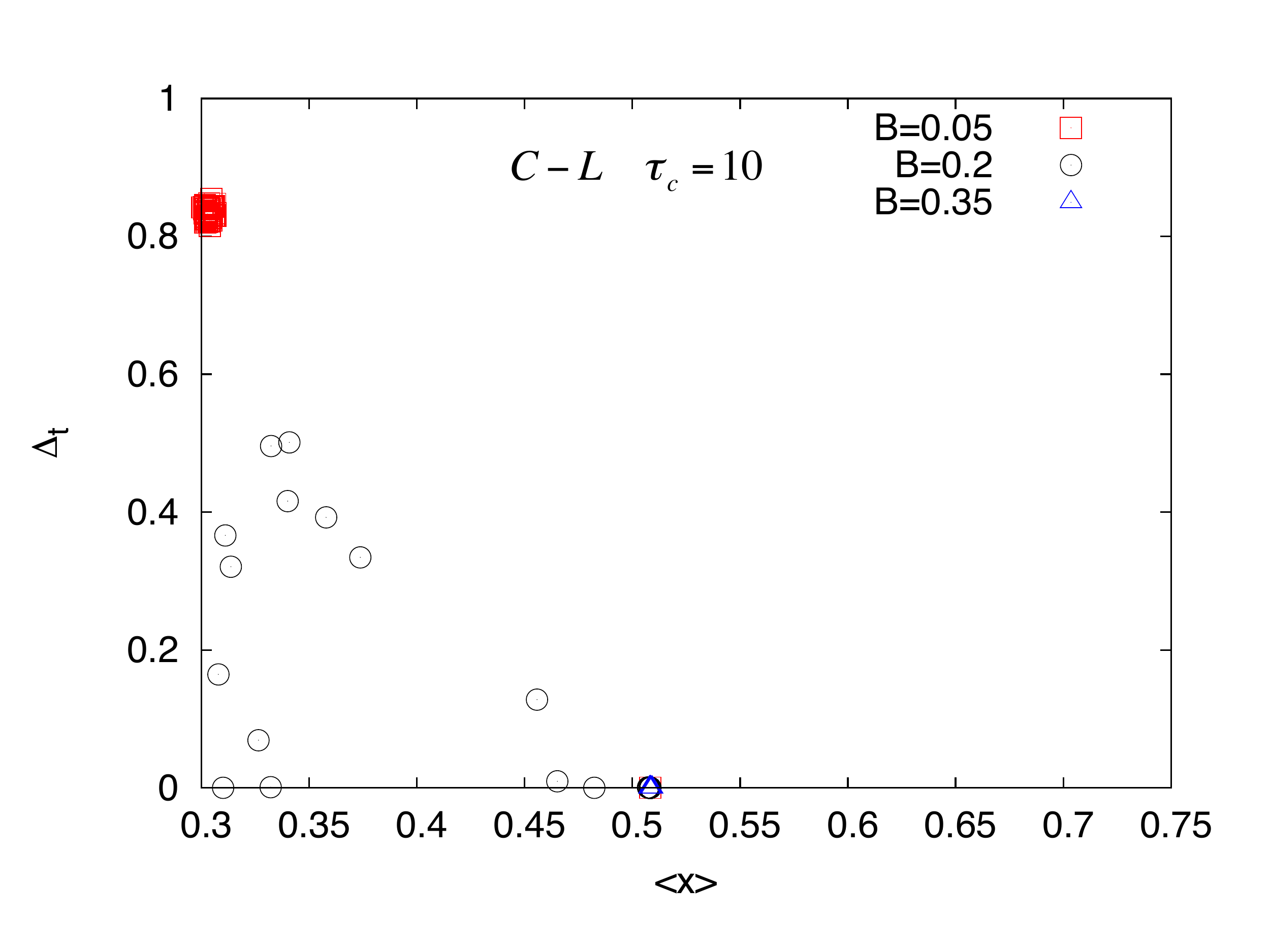}
}
\subfigure[]
{
\label{B}
\includegraphics[width=0.43\textwidth]{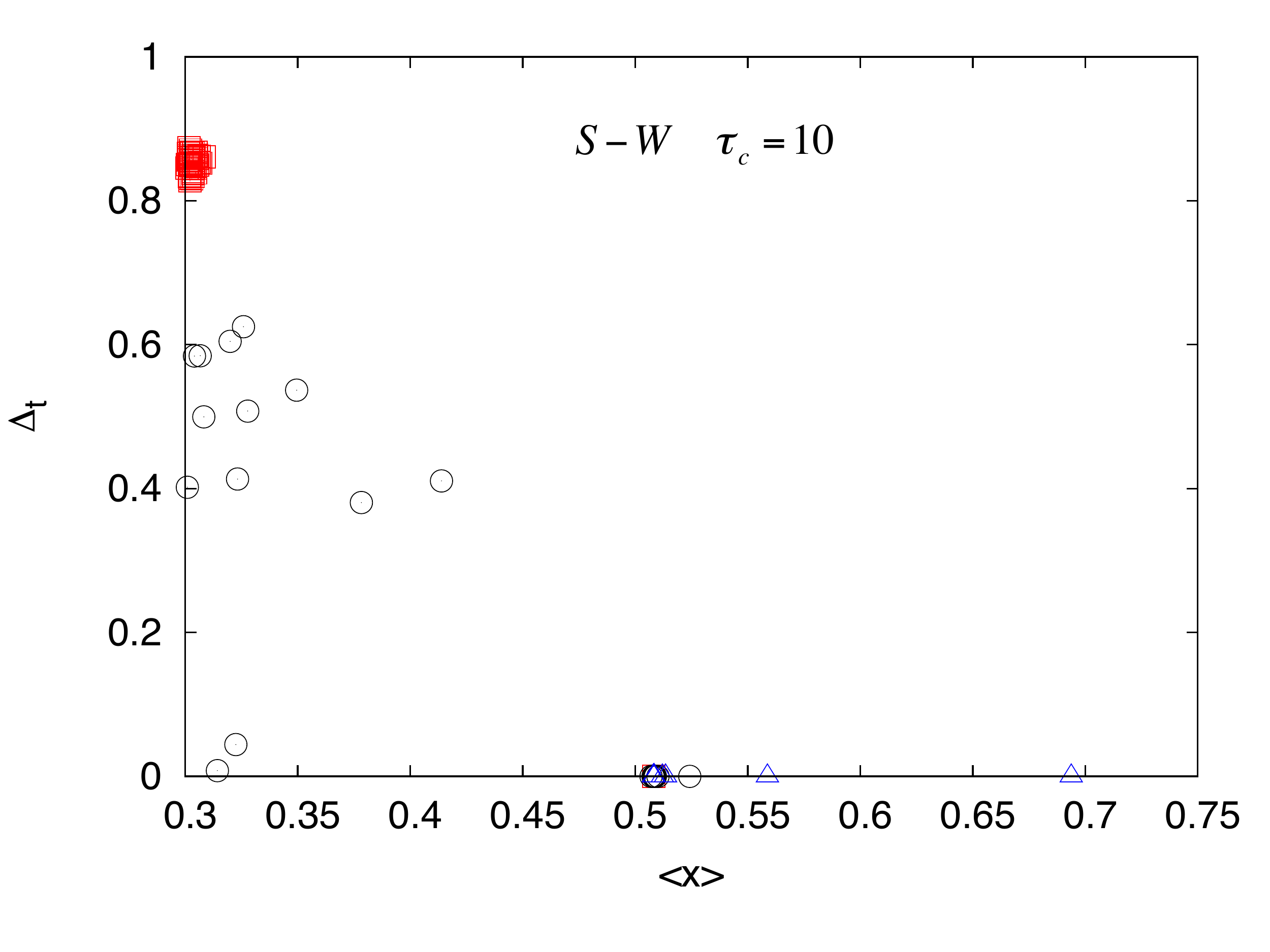}
}
\end{center}
\caption{Spatially white but temporally colored perturbations of parameter $\delta$: scatterplot $(\Delta_t,\langle x \rangle)$. Panel \subref{A}: Cai-Lin noise;  panel \subref{B}: sine-Wiener noise. In both cases: $\tau_c=10 \; s$. Number of simulated systems: $50$.
}
\label{fig_2}
\end{figure}

\begin{figure}[t]
\begin{center}
\subfigure[]
{
\label{A}
\includegraphics[width=0.3\textwidth, angle=-90]{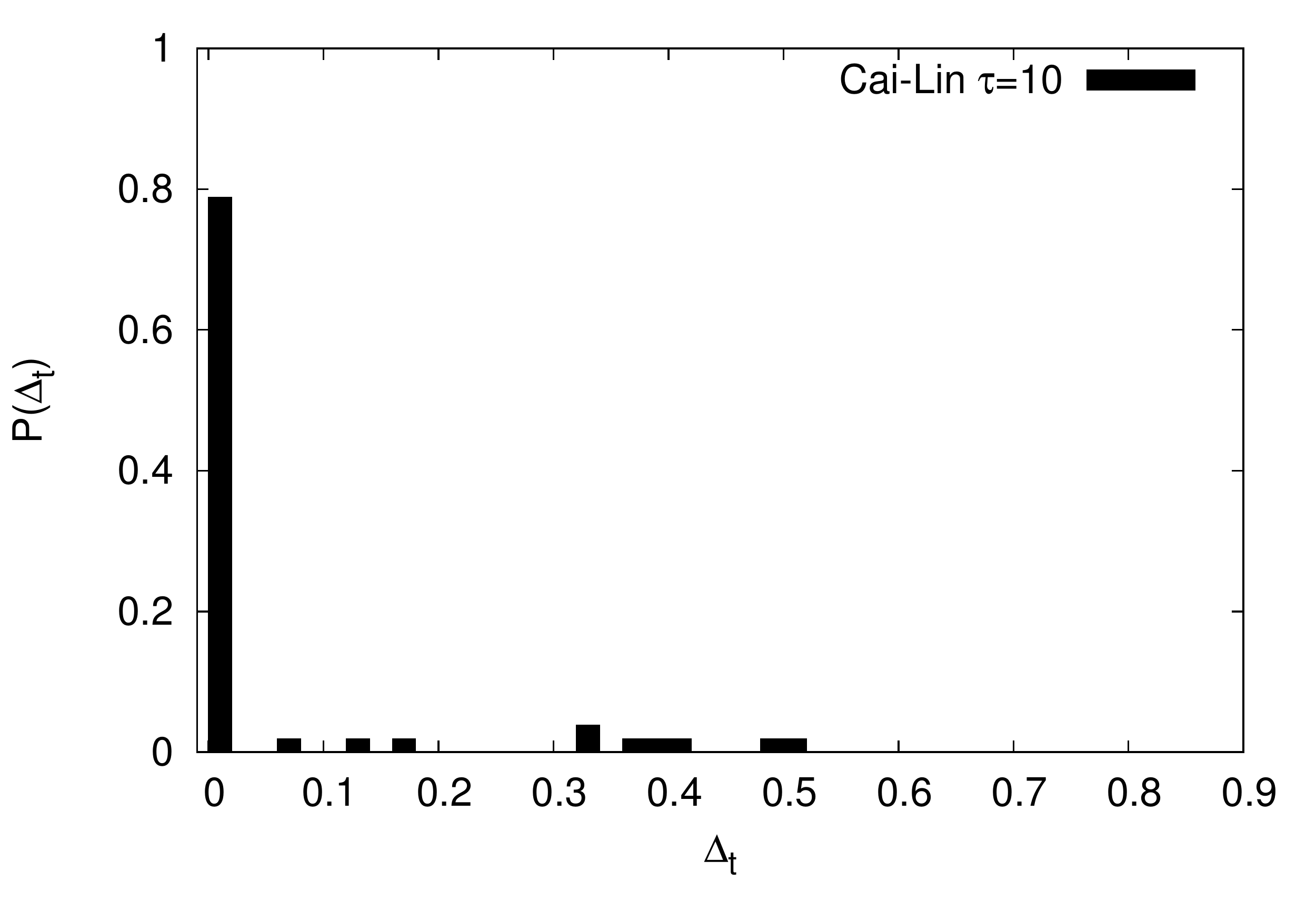}
}
\subfigure[]
{
\label{B}
\includegraphics[width=0.3\textwidth, angle=-90]{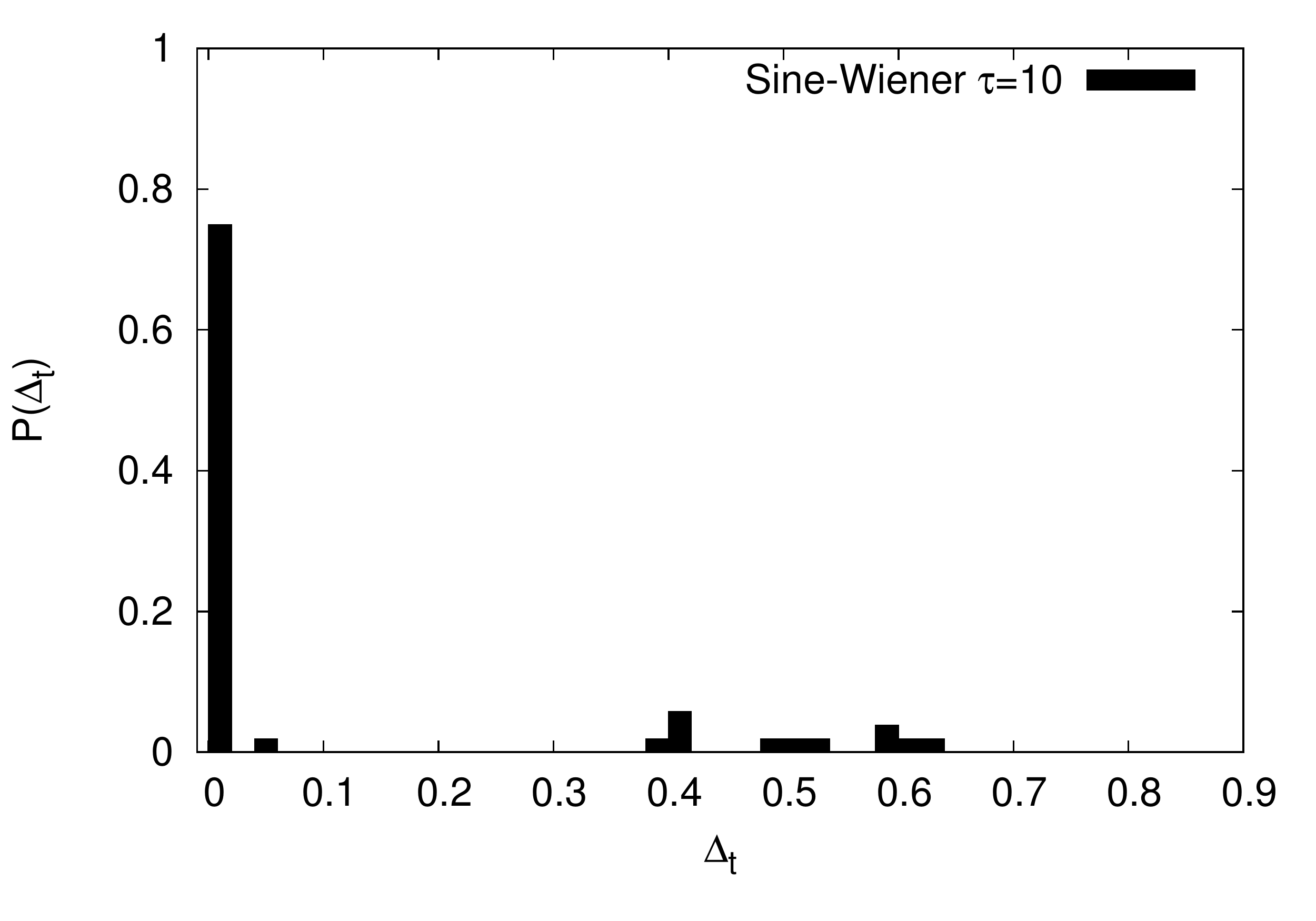}
}
\subfigure[]
{
\label{C}
\includegraphics[width=0.3\textwidth, angle=-90]{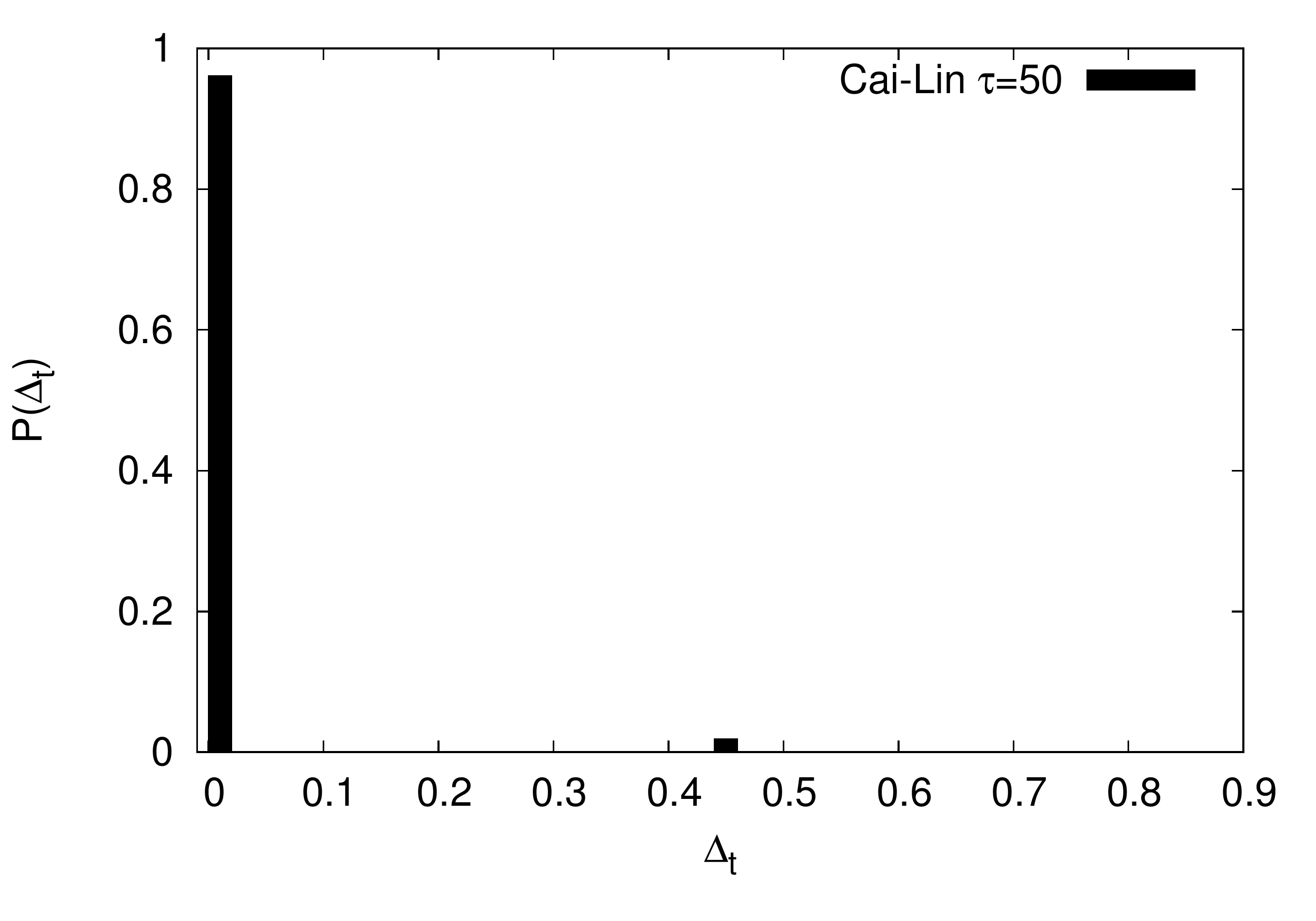}
}
\subfigure[]
{
\label{D}
\includegraphics[width=0.3\textwidth, angle=-90]{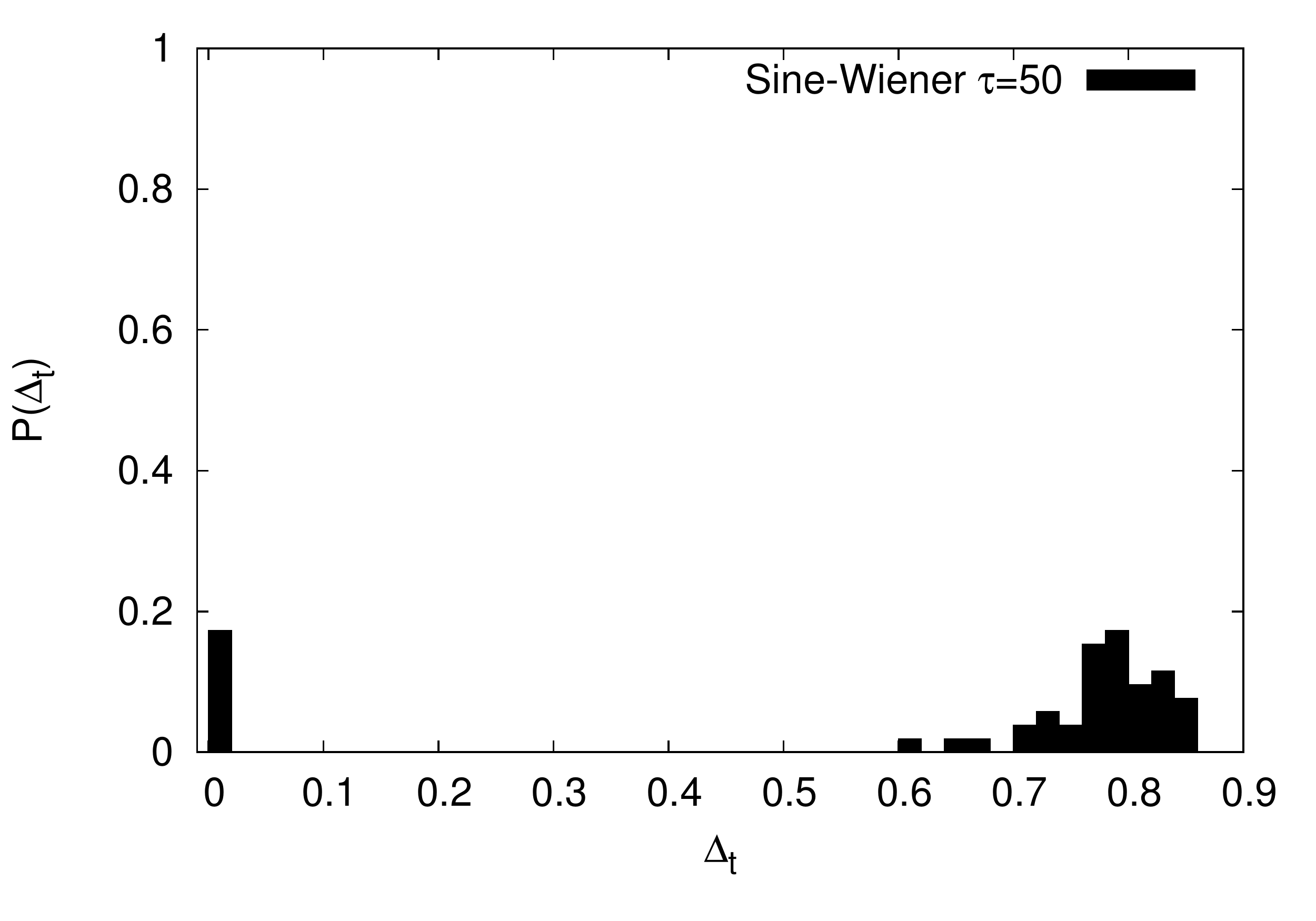}
}
\end{center}
\caption{Spatially white but temporally colored perturbations of parameter $\delta$: heuristic histograms of the distributions $P(\Delta_t)$. Upper panels: $\tau_c=10 \; s$; lower panels: $\tau_c =50 \; s$. Panels \subref{A} and \subref{C}: response to Cai-Lin noise; panels \subref{B} and \subref{D}: response to sine-Wiener noise. In all panels: $B=0.2$.
The response to increasing $\tau_c$ is different for the two kinds of noises: the depolarization probability increases for Cai-Lin noise, decreases for sine-Wiener noise.
}
\label{fig_4}
\end{figure}

In such a case, we observed that the response to the noise strongly depends on the type of the perturbation. Indeed, setting $\tau_c =10 \; s$ and $B=0.2$, both Cai-Lin and sine-Wiener noises cause in many cases depolarization of the cell. This is illustrated by the scatterplots shown in figure \ref{fig_2}, and in the distribution of $\Delta_t$ shown in the upper panels of figure \ref{fig_4}.

\begin{figure}[t]
\begin{center}
\subfigure[]
{
\label{A}
\includegraphics[width=0.45\textwidth]{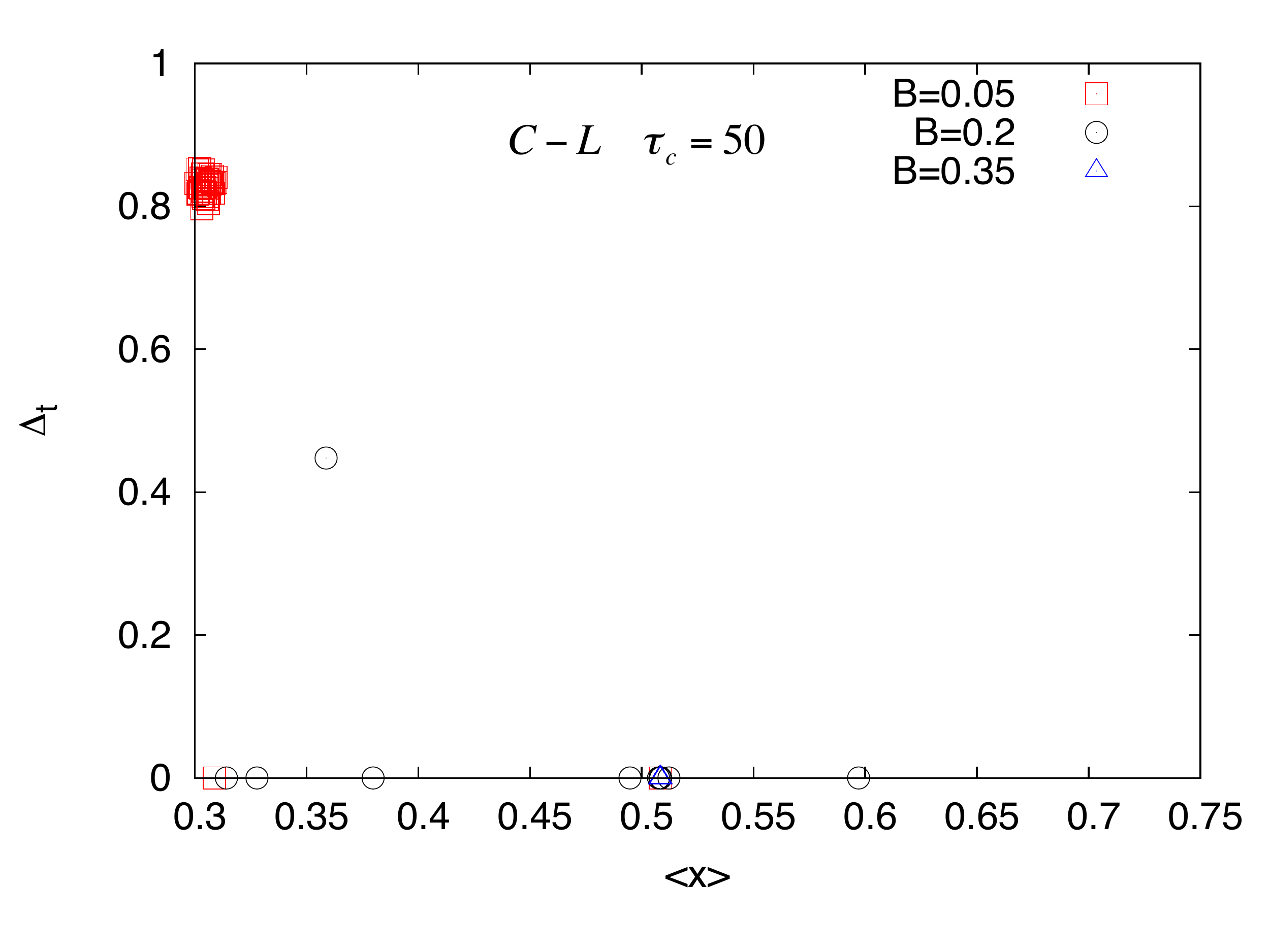}
}
\subfigure[]
{
\label{B}
\includegraphics[width=0.45\textwidth]{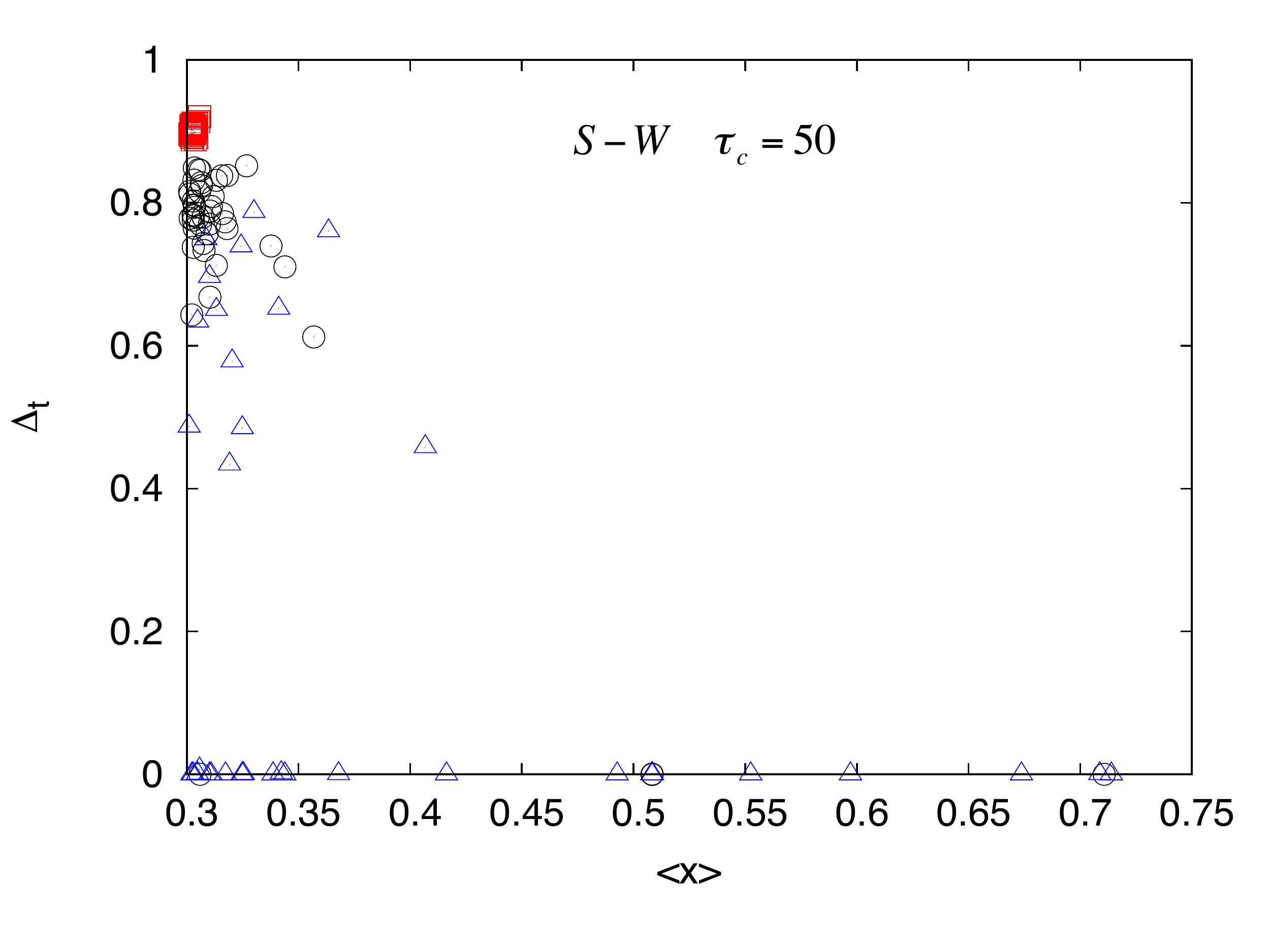}
}
\end{center}
\caption{Spatially white but temporally colored perturbations of parameter $\delta$: scatterplot $(\Delta_t,\langle x \rangle)$. Panel \subref{A}: Cai-Lin noise;  panel \subref{B}: sine-Wiener noise. In both cases: $\tau_c=50 \; s$, $B= 0.2$. Number of simulated systems: $50$.
}
\label{fig_3}
\end{figure}

By increasing the temporal autocorrelation $\tau_c$ up to $50 \; s$ we observed two dichotomic behaviors. If the perturbation is a Cai-Lin noise then the increase of  $\tau_c$ causes a larger number of depolarizations (see figures \ref{fig_3}a and \ref{fig_4}c). On the contrary, if the bounded noise is of sine-Wiener type then one observes a larger probability that the cell maintains the polarization induced by the external deterministic cue (see figures \ref{fig_3}b and \ref{fig_4}d).

As far as perturbations in $\omega$ and $\gamma$ are concerned, fluctuations in $\gamma$ induce effects comparable to the ones caused by noises affecting $\delta$, whereas perturbations of $\omega$ depolarize to a lesser extent. Note that the perturbation of $\omega$ may summarize the effect of extrinsic noise in the feedback mechanism.

\begin{figure}[t]
\begin{center}
\subfigure[]
{
\label{A}
\includegraphics[width=0.3\textwidth, angle=-90]{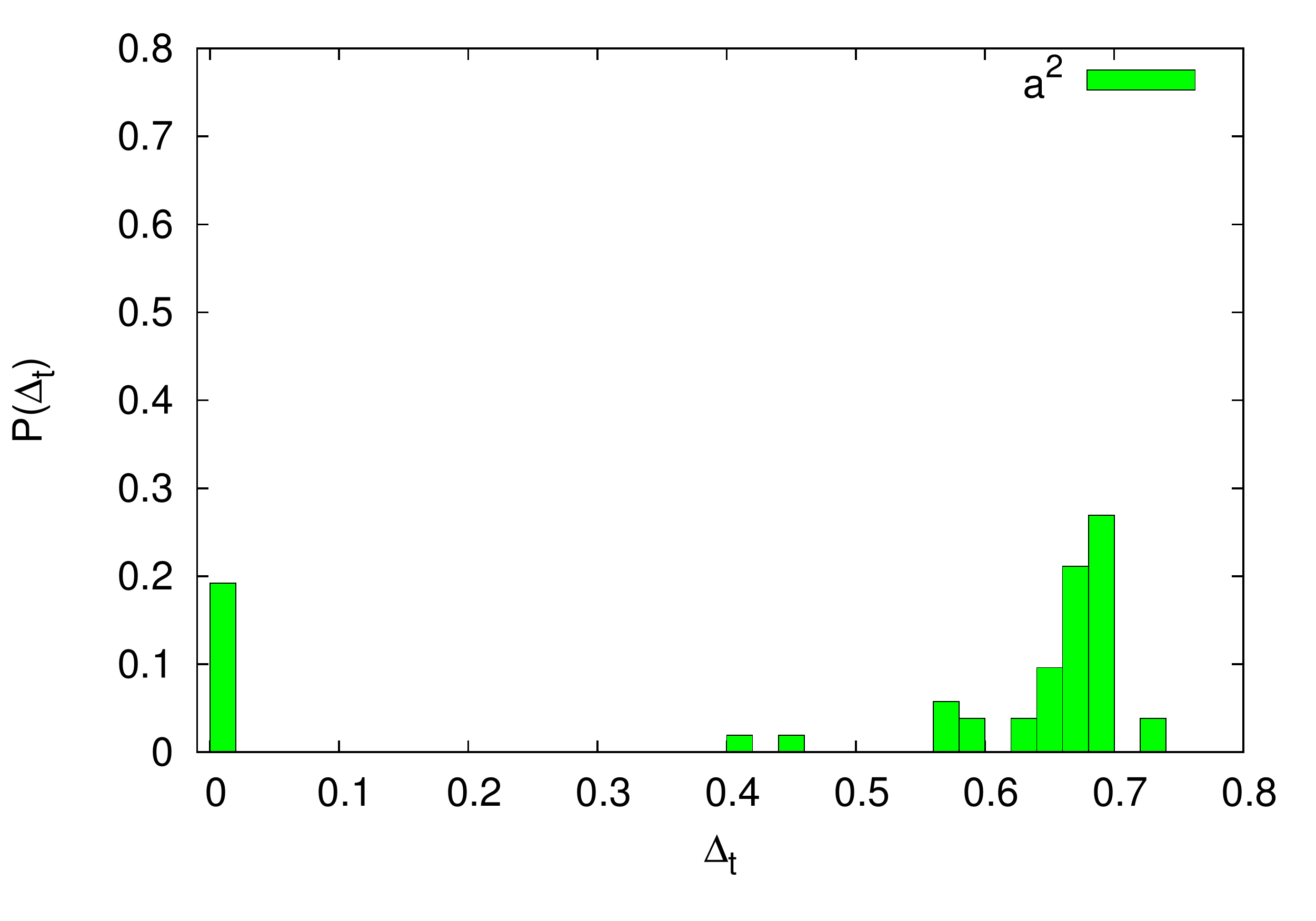}
}
\subfigure[]
{
\label{B}
\includegraphics[width=0.3\textwidth, angle=-90]{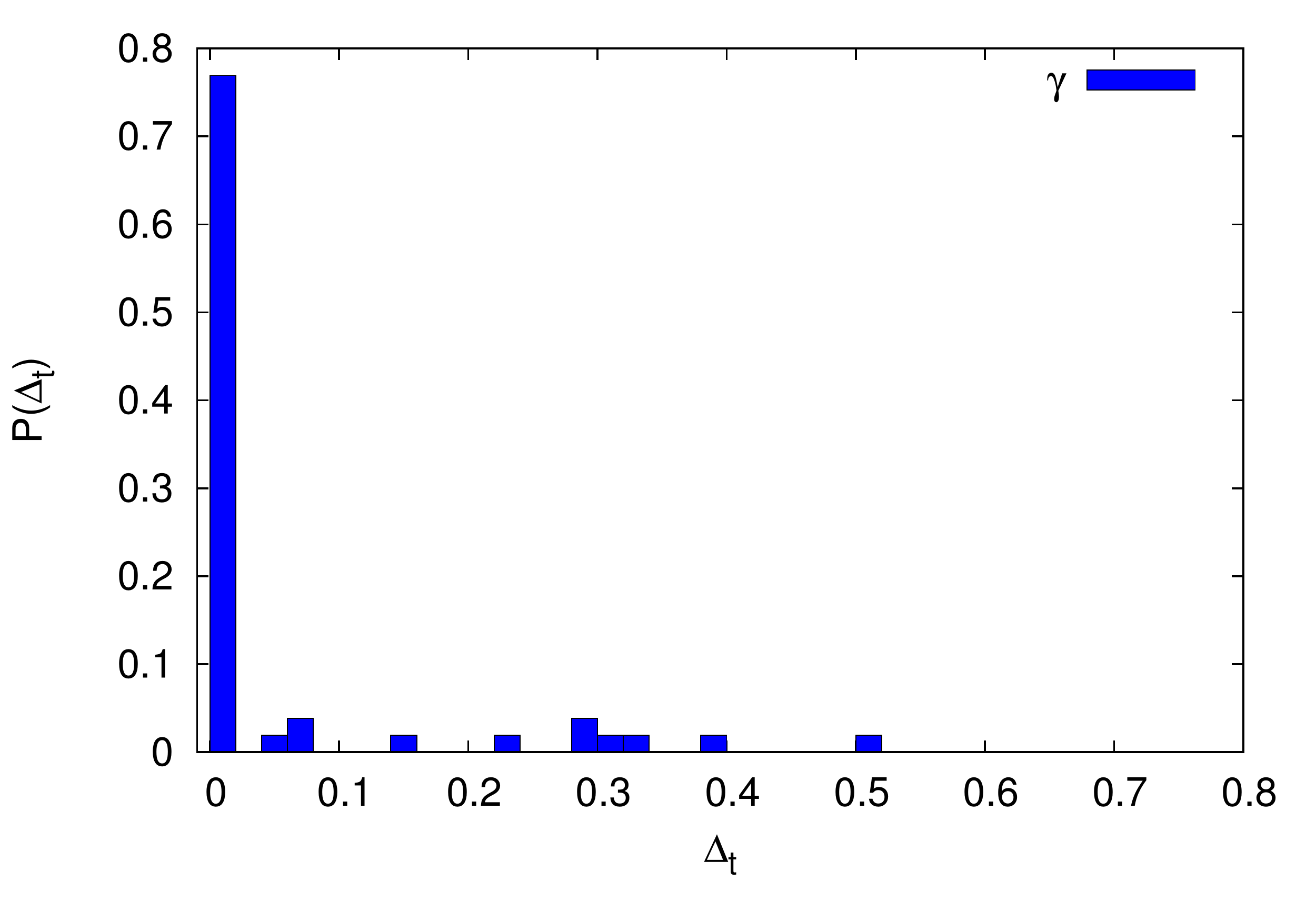}
}

\end{center}
\caption{
Spatially white but temporally colored perturbations affecting $\omega$ (panel \subref{A}) and $\gamma$ (panel \subref{B}): heuristic histograms of the distributions $P(\Delta_t)$. In both cases: Cai-Lin noise with $\tau_c =10 \; s$ and $B=0.2$.
}
\label{fig_5}
\end{figure}

\begin{figure}[t]
\begin{center}
\includegraphics[width=0.5\textwidth, angle=-90]{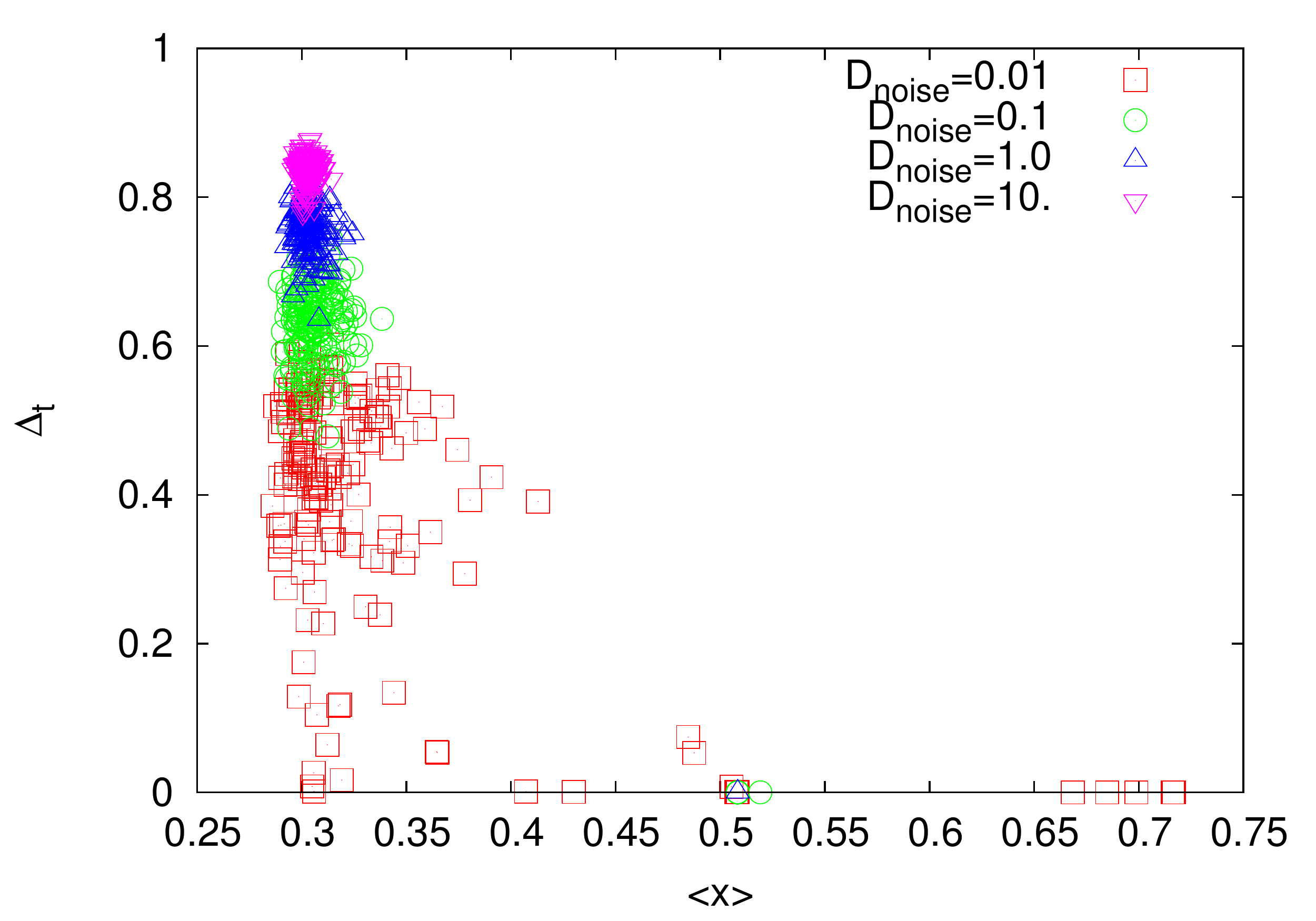}
\end{center}
\caption{Spatially and temporally colored noise: scatterplot  $(\Delta_t, \langle x \rangle)$ for various values of $D_{noise}$. Here the perturbed parameter is  $\gamma$. Type of noise: Cai-Lin. Parameters $\tau_c =10 \; s$, $B = 0.2$. Number of points for each series is $200$.
}
\label{fig_6}
\end{figure}

\begin{figure}[t]
\begin{center}
\subfigure[]
{
\label{A}
\includegraphics[width=0.3\textwidth, angle=-90]{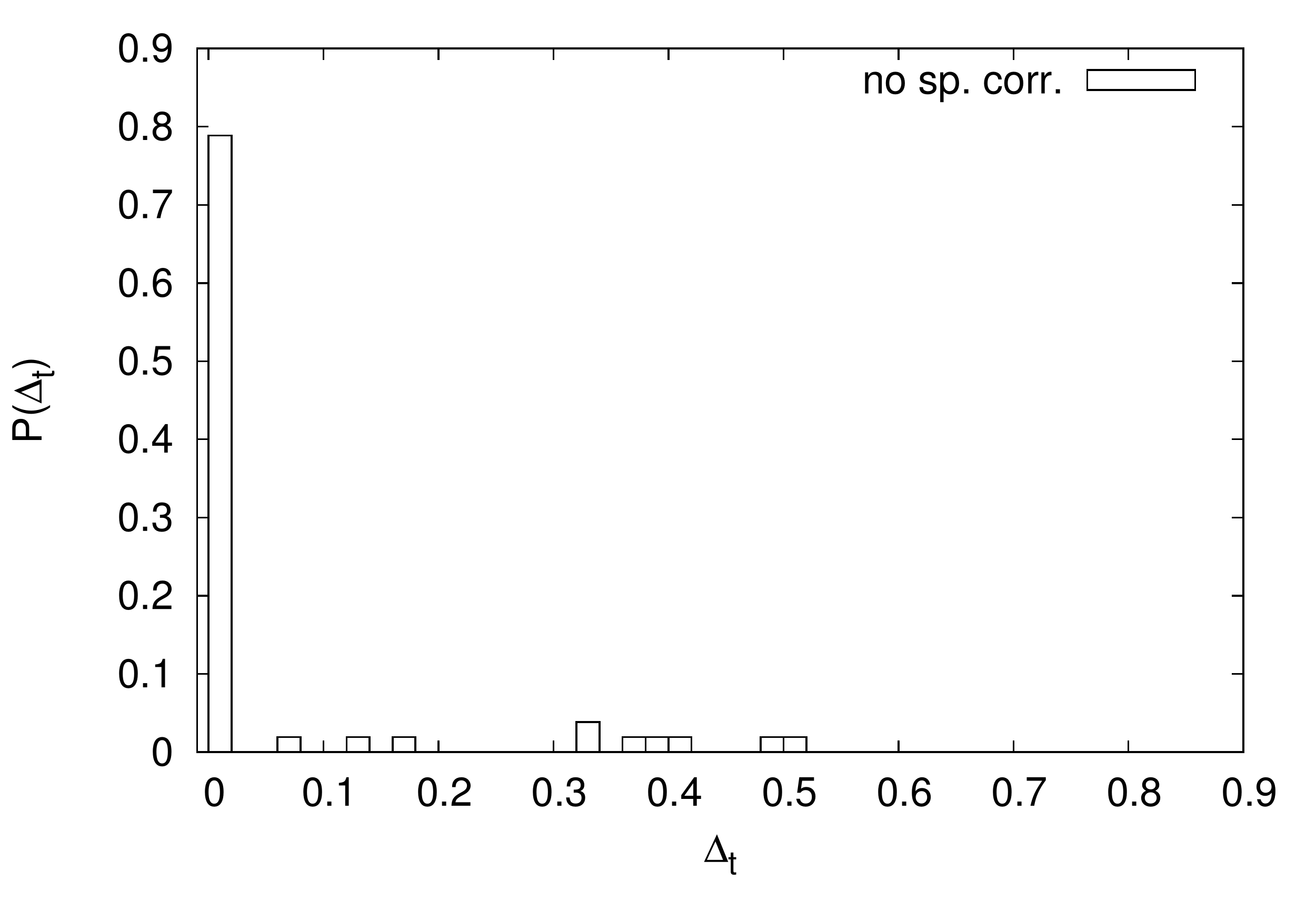}
}
\subfigure[]
{
\label{B}
\includegraphics[width=0.3\textwidth, angle=-90]{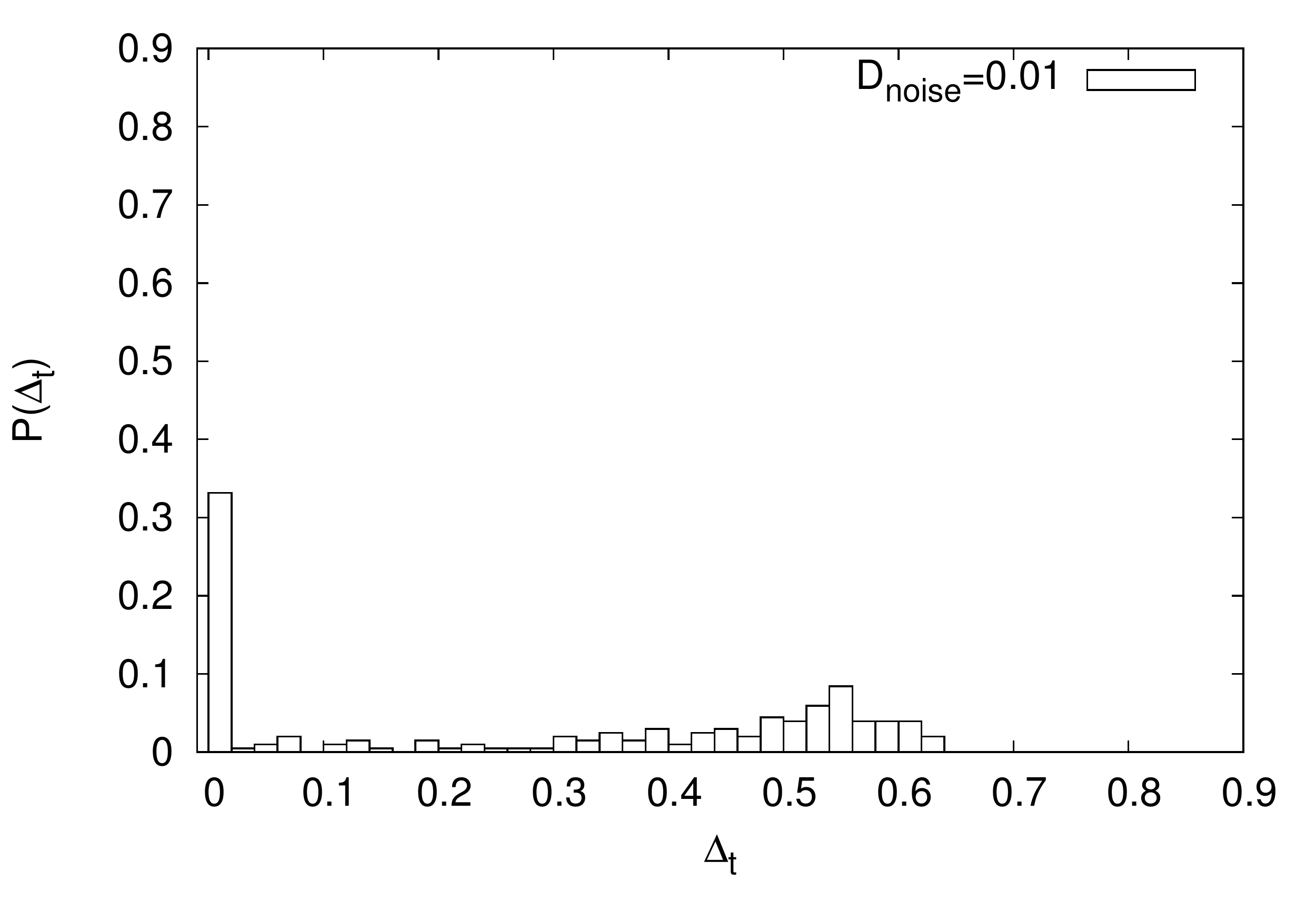}
}
\subfigure[]
{
\label{C}
\includegraphics[width=0.3\textwidth, angle=-90]{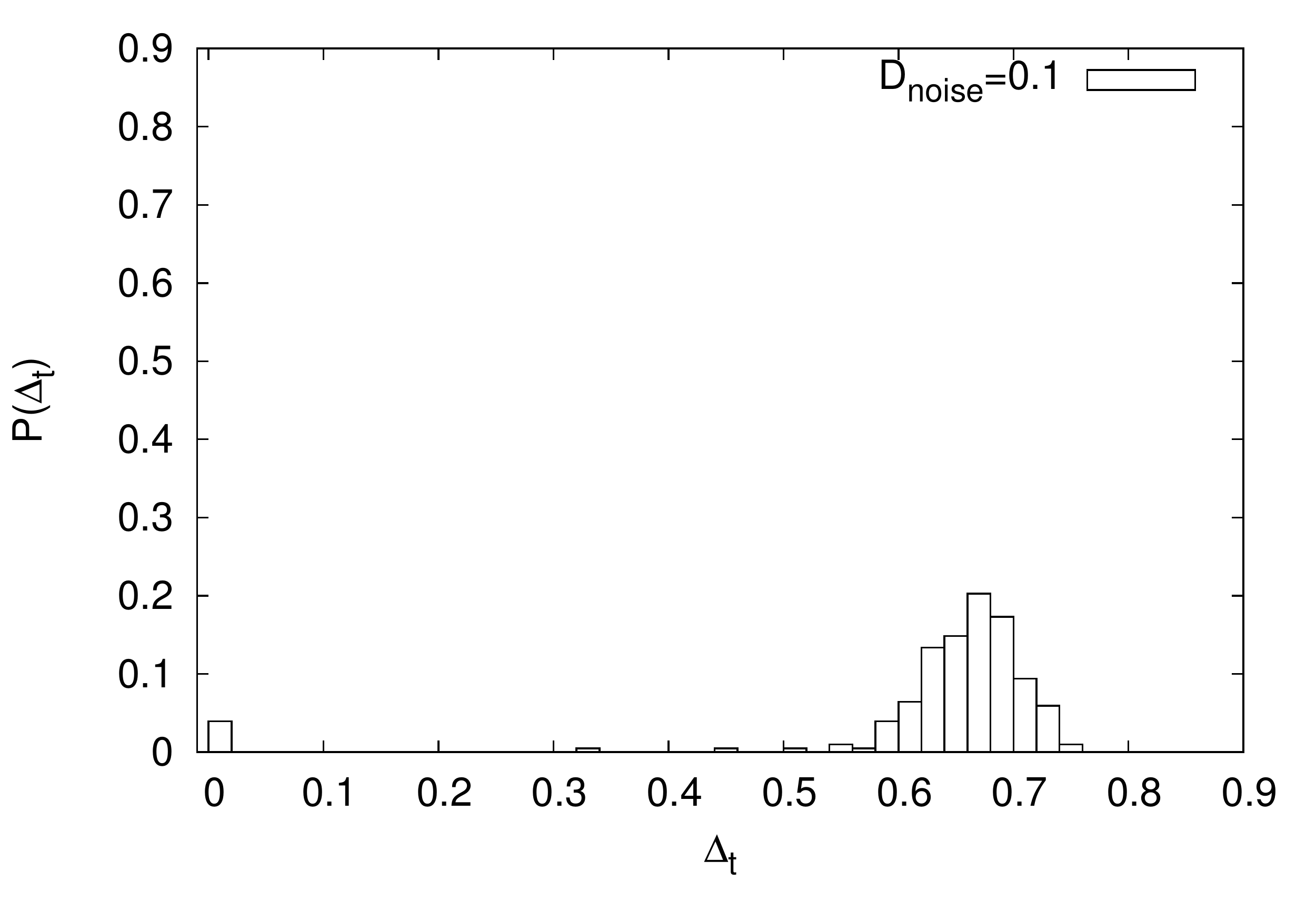}
}
\subfigure[]
{
\label{D}
\includegraphics[width=0.3\textwidth, angle=-90]{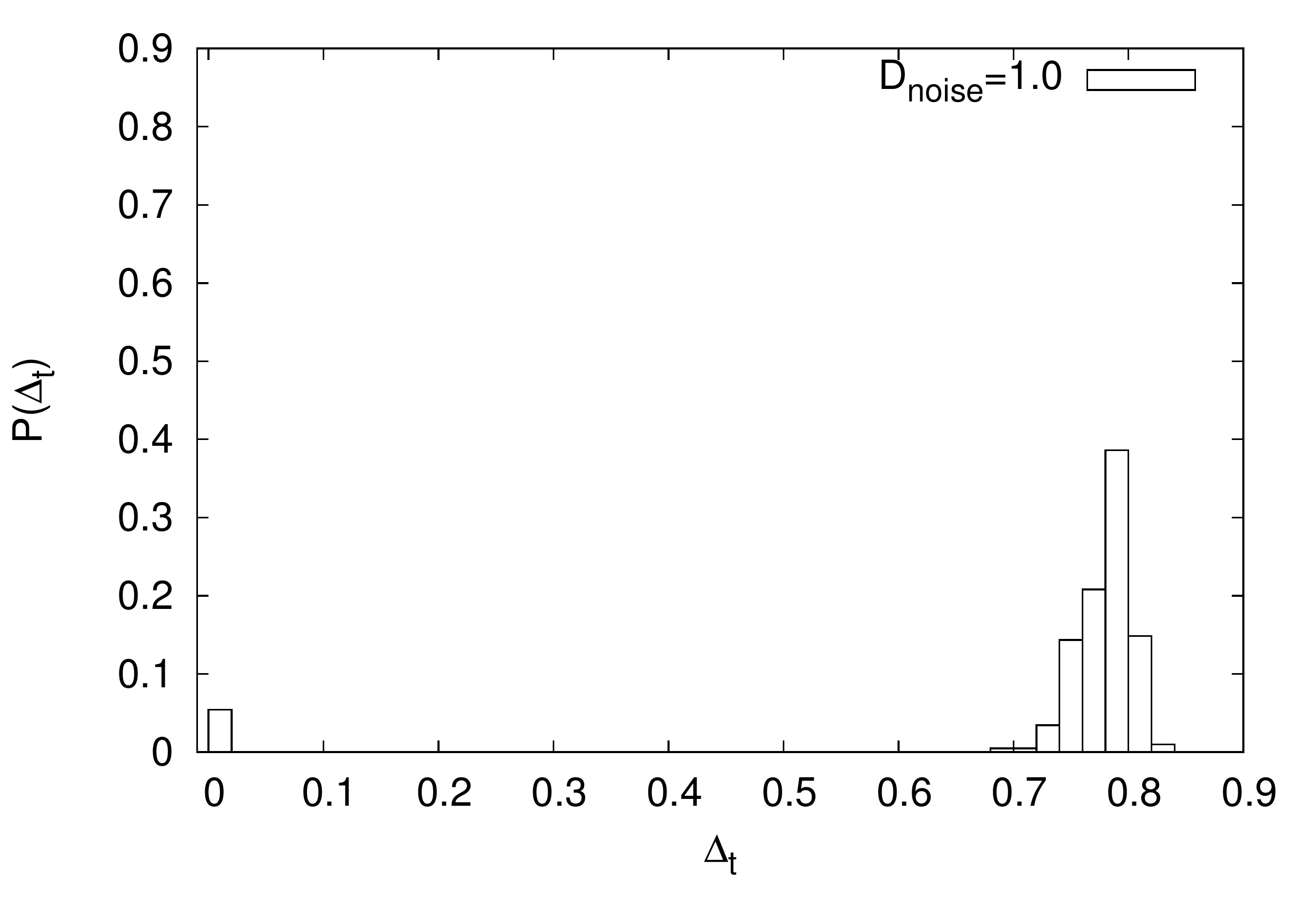}
}
\subfigure[]
{
\label{E}
\includegraphics[width=0.3\textwidth, angle=-90]{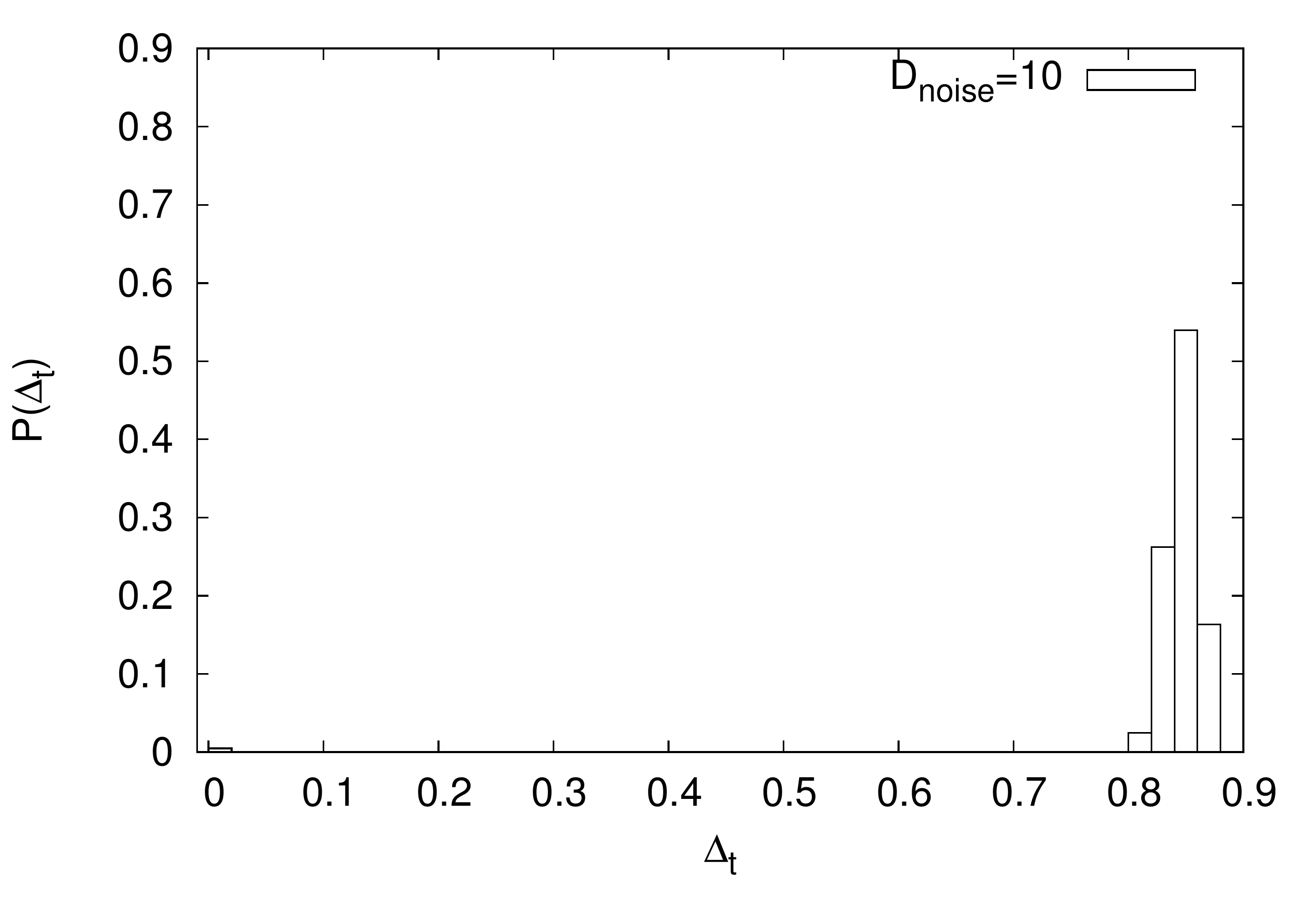}
}
\subfigure[]
{
\label{F}
\includegraphics[width=0.3\textwidth, angle=-90]{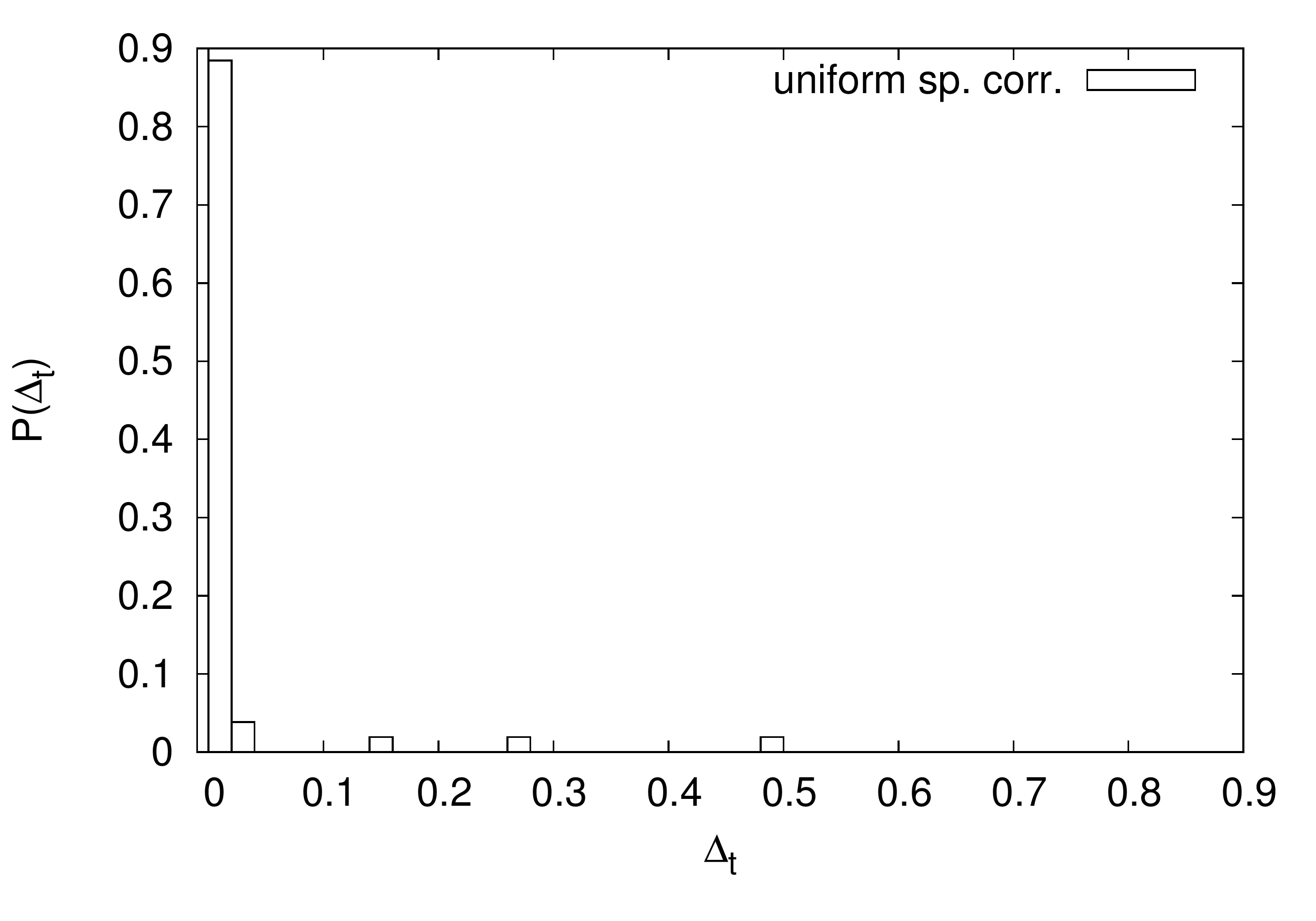}
}
\end{center}
\caption{Histograms of $P(\Delta_t)$ in case of various degrees of spatial correlation of a Cai-Lin perturbation applied to parameter $\delta$. Panel \subref{A}: $D_{noise} =0$; panel \subref{B}: $D_{noise} =0.01$; panel \subref{C}: $D_{noise} =0.1$; panel \subref{D}: $D_{noise} =1$;
panel \subref{E}: $D_{noise} =10$; panel \subref{F}: spatially uniform noise. Other parameters: $\tau_c =10 \; s$, $B=0.2$. 
}
\label{fig_8}
\end{figure}

We now examine the effects of spatially correlated noises. In figures \ref{fig_6}-\ref{fig_8} we show that the presence of non-null spatial correlation, i.e. $D_{noise}>0$, increases not only the probability of maintaining the cell polarization, but also its intensity (i.e. the magnitude of spatial gradient of $a$). Indeed, in such cases simulations suggest not only a decrease of the probability of observing small or null $\Delta_t$, but also a significant increase of the probability that $\Delta_t$ is 'large'. For example: if $D_{noise} = 0.01$ the probability $P(0)$ is quite large (in $30\%$ of cases the cell is not polarized), whereas for $D_{noise}\ge 0.1$ $P(0)$ is small and it is  a decreasing function of $D_{noise}$.

We also simulated the response of the WPP model to a noise that is spatially uniform but temporally varying. Namely: \textit{i)} instead of the spatio-temporal Cai-Lin noise, we used a noise $\beta(x,t) = \xi(t)$, where $\xi(t)$ is the temporal Cai-Lin noise; \textit{ii)} instead of the sine-Wiener noise we employed a noise $\beta(x,t)= B sin(2 \pi \xi(t) )$ where $\xi(t)$ si the OU noise. Since in this case the spatial correlation length is infinite, we expected to observe that all the simulated cells could maintain their polarization. Quite interestingly, we observed that this spatially uniform 
noise induces (see fig. \ref{fig_8}f) in the model a behavior that is comparable to the one induced by spatially white noises (compare with previous fig. \ref{fig_8}a).

\FloatBarrier

\subsection{Initial random state}
Here we illustrate the results of simulations performed without adding the external deterministic cue, but considering an initial random distribution of $A$ , following the equation (\ref{rcue}) with $R=3$.

In absence of noise, we observed three possible kinds of steady states:  polarization (left or right, in this case it does not matter) or a central 'hump'. 

\begin{figure}[t]
\begin{center}
\includegraphics[width=0.5\textwidth, angle=-90]{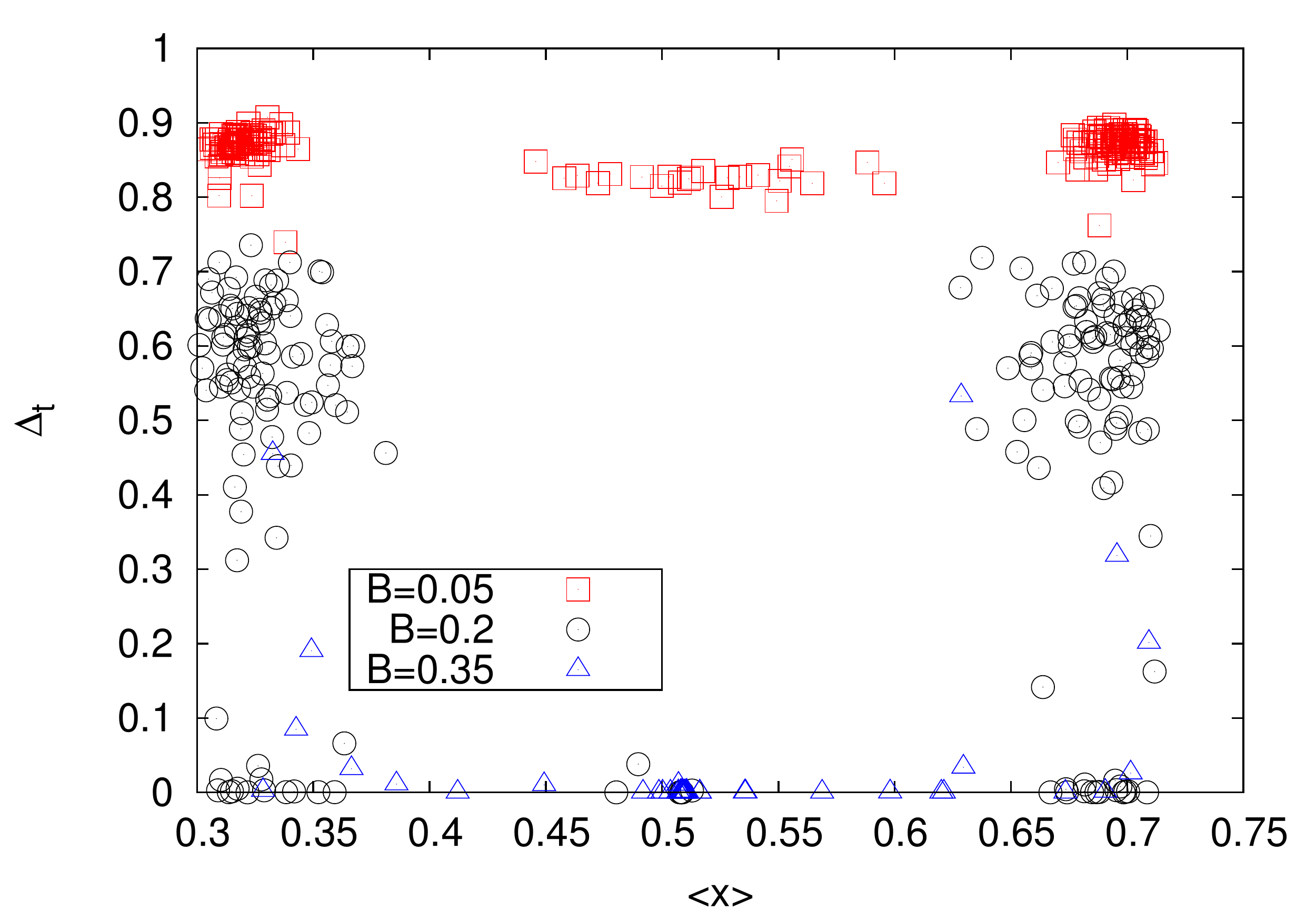}
\end{center}
\caption{
Random initial conditions and spatially white sine-Wiener perturbation of $\delta$. Scatterplot  $(\Delta_t,\langle x \rangle)$ for three values of $B$. Other parameters: $\tau_c = 10 \; s$.
}
\label{ciccio}
\end{figure}

\begin{figure}[t]
\begin{center}
\includegraphics[width=0.9\textwidth]{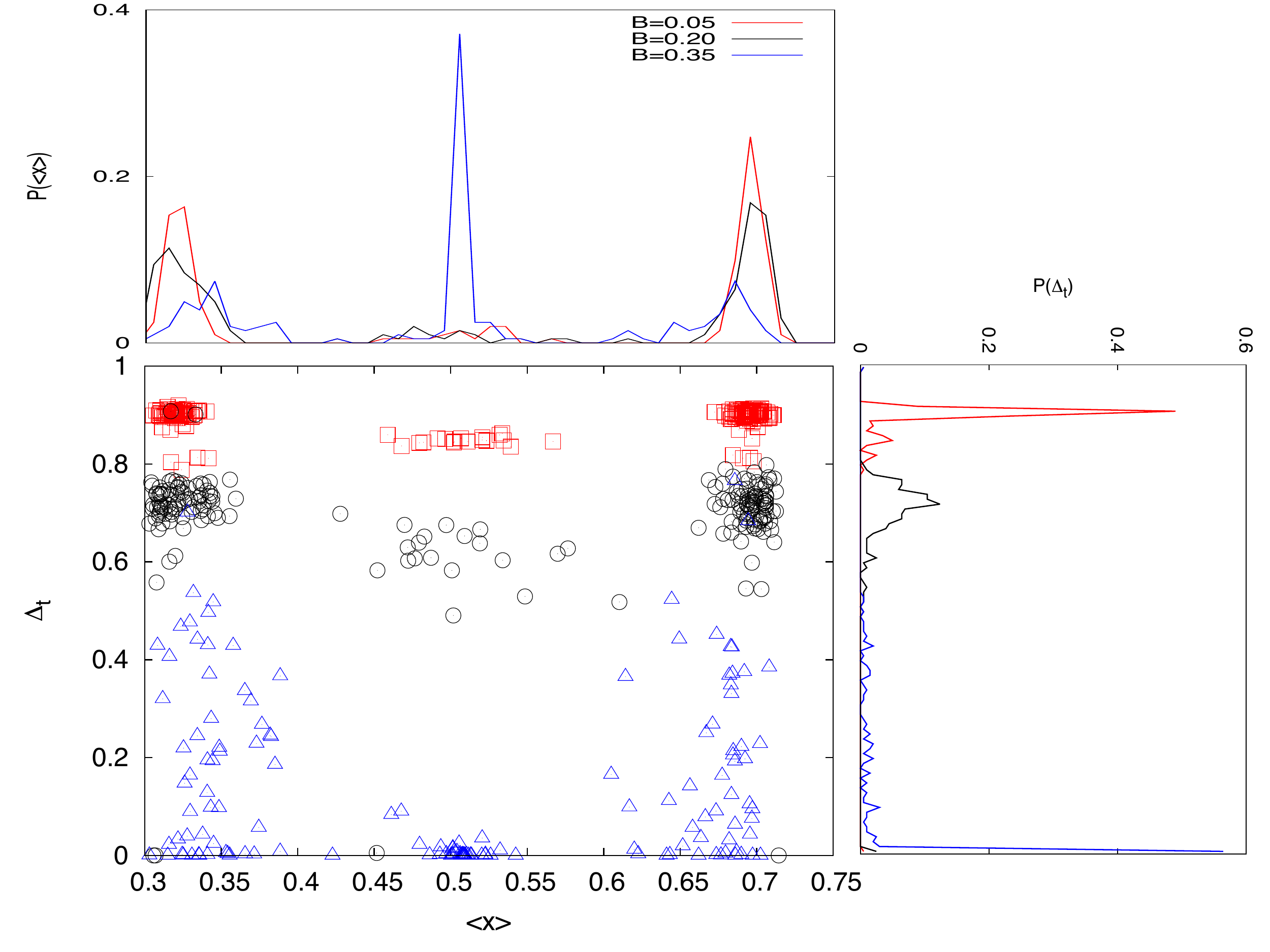}
\end{center}
\caption{Random initial conditions and spatially white Cai-Lin perturbation of $\omega$. Scatterplot  $(\Delta_t,\langle x \rangle)$ and corresponding histograms of the distributions $P(\Delta_t)$ and $P(\langle x \rangle)$. Other parameters: $\tau_c=10 \; s$. 
}
\label{pluto}
\end{figure}

As far as spatially white noises are concerned, we observed a behavior depending on the perturbed parameter, but not on the type of noise. Indeed:
\begin{itemize}
\item in case of perturbation of $\delta$ (see figure \ref{ciccio}, where the noise is of sine-Wiener type): for $B=0.2$ solutions with central hump are not observed, and in some cases cell is not polarized. Absence of polarization is always observed if  $B=0.35$;
\item in case of perturbation of $\omega$ (see figure \ref{pluto}) or of $\gamma$ (not shown): for $B=0.2$ cells seldom is not polarized and the number of solutions with central 'hump' is roughly similar to that observed if $B=0.05$. For $B=0.35$ in sometime cases the cells remain polarized (and no cells has the 'hump' pattern).
\end{itemize}

Finally, also here the increase of spatial correlation restores the polarization.

\section{Concluding remarks}
Our main aim in this work was to study the robustness to extrinsic stochasticity of an interesting mechanism explaining both cued and un-cued cellular polarization: the wave-pinning \cite{k2008}.

We modeled the above-mentioned extrinsic perturbations by means of  spatio-temporal bounded noises that model perturbations in a more realistic way than Gaussian noises, which are unbounded. In particular, we adopted the following two noises: the Cai-Lin \cite{deFradOnpre} and the sine-Wiener \cite{deFranDOnoSW} noises, which we recently proposed. 

The robustness of the WPP model to extrinsic noise was, in particular, investigated by means of suitable statistics summarizing the polarization probability, the fluctuations and the intensity of the polarization.

An interesting feature of the deterministic WPP model is that it is deterministically robust, in the sense that  most models exhibiting the wave-pinning phenomenon are such that the waves 'freezes' for a specific isolated value of a given parameter \cite{Sepulchre00,k2008}. In the WPP model, instead, the system self regulates and the wave freezes for an entire range of the chosen bifurcation parameter\cite{k2008}.

Here we showed that under extrinsic stochastic perturbations, the maintenance or loss of polarization strongly depends on both spatial and temporal colors, as well as on the specific kind of noise.

Indeed, in case of Cai-Lin noise the increase of temporal correlation induces a decrease of polarization probability, whereas in case of sine-Wiener noise one observes the opposite phenomenon.

Moreover, in both cases passing from white to spatially correlated noise induces a loss of polarization. However, in the simulations where we employed a spatially constant noise we again obtained a large probability of polarization.

In other words,  the WPP model exhibits a sort of contrast of colors (the spatial and the temporal) and of noise types.

Concerning the possible use of the WPP model to describe the spontaneous cellular polarization emerging from random initial conditions, beside the polarized and global oscillating states, we observed (in line with \cite{k2008}) the emergence of "non-polar" patterns that are characterized by a hump in the density of A, located in the center of the cell. Moreover, the effect of the amplitude of the noise  (and of the spatial coupling) is 'non-monotone'. Indeed, it is possible to find an intermediate amplitude of the noise (or spatial coupling strength) by which both the probabilities of unpolarized oscillating states and of 'humped' states are low.

Summarizing the above observations, we may say that the answer to the main substantive question 'Is the WPP model robust?' is: 'it depends on the context where the cell is embedded'. In other words, being equal the amplitude of the external perturbations, the robustness of the wave-pinning-based mechanism of cellular polarity strongly depends on the kind of extrinsic stochasticity that affects the WPP network. Note also that the ability of this mechanism to describe a spontaneous cellular polarization is questionable (as also stressed in \cite{KeshetPLOS11}) because of the presence of humped solutions. Although in presence of extrinsic noise the hump may disappear, however in those cases the onset of noise-induced spatially homogenous states is observed.

As far as the onset of wave-pinning-induced polarization in case of low number of molecules, studied in \cite{Keshet2012}, we remind the reader that for moderate number of molecules a stochastic reaction system may be approximated by the corresponding deterministic system plus suitable multiplicative external white Gaussian noises. This is a well-known result in statistical and chemical physics \cite{gammaitoni, Vanka90}, also known in Systems Biology as 'chemical Langevin equation' method \cite{Gillespie00}. Thus here we might be tempted to read the result of paper \cite{Keshet2012} as fragility of the WPP mechanism to white noise perturbations. However, we stress here that for computational reasons the simulations of \cite{Keshet2012} has been performed by assuming that the number of molecules of the species $B$ is spatially homogeneous. Moreover we recall that it is the equation (\ref{wppb}) that guarantees the robustness of WPP model. This might imply that the stochastic version of WPP model could be more robust than it has been reported in the numerical simulations of \cite{Keshet2012}.

Thus, our aim is to explore in a future work the full discrete stochastic version of equations (\ref{wppa})-(\ref{wppb}), also including extrinsic noises. Indeed, the interplay of intrinsic and bounded extrinsic noise may be of interest in Systems Biology, as stressed in \cite{GiulioDon12PONE12}.

\section*{Acknowledgment}
This work was performed in the framework of the Integrated Project “P-medicine" from data
sharing and integration via VPH models to personalized medicine” (project ID: 270089), which is
partially funded by the European Commission under the 7th framework program.

\bibliographystyle{spphys} 
\bibliography{Bibliography} 
\end{document}